\documentclass[12pt,letterpaper]{article}
\usepackage{savesym}
\usepackage{pdflscape}
\usepackage{wrapfig}
\usepackage{amssymb}
\usepackage{amsthm}
\usepackage{amsmath}
\usepackage{graphicx}
\usepackage[usenames, table]{xcolor}
\usepackage{verbatim}
\usepackage{times}
\usepackage[titles]{tocloft}
\usepackage{enumitem}
\usepackage{bm}
\usepackage{sidecap}
\usepackage{epstopdf}
\usepackage{pdfpages}
\usepackage{hyphenat}
\usepackage[colorinlistoftodos]{todonotes}
\usepackage{soul}
\usepackage{multicol}
\usepackage{float}
\usepackage{multibib}
\usepackage{geometry}
\usepackage[small,compact]{titlesec}
\usepackage[normal,labelsep=colon,skip=2pt]{caption}
\usepackage{setspace}
\usepackage{subfig}
\usepackage{tocloft}
\usepackage{url}
\usepackage{floatflt}
\usepackage{array}
\usepackage{wasysym}
\usepackage{pgfgantt}
\usepackage[colorlinks=true,
            linkcolor=black,
            urlcolor=red,
            citecolor=blue]{hyperref}

\usepackage{cite}
\usepackage{lineno}
\usepackage[title,titletoc]{appendix}
\usepackage[affil-it]{authblk}

\newcommand*{\TitleFont}{%
      \usefont{\encodingdefault}{\rmdefault}{b}{n}%
      \fontsize{16}{20}%
      \selectfont}
\graphicspath{{./figs/}{../figs/}{../}}
\savesymbol{comment} 
\usepackage{changes}
\definechangesauthor[color=purple]{xl}
\definechangesauthor[color=orange]{hx}

\definecolor{lightblue}{rgb}{.90,.95,1}
\definecolor{darkgreen}{rgb}{0,.5,0.5}

\begin{document}


\title{\TitleFont Determining \textit{ a priori} a RANS model's applicable range via global epistemic uncertainty quantification}

\author[1]{Xinyi Huang}
\author[1]{Naman Jain}
\author[2]{Mahdi Abkar}
\author[1]{Robert Kunz}
\author[1]{Xiang Yang}

\date{\vspace{-5ex}}

\affil[1]{Mechanical Engineering, Pennsylvania State University, Pennsylvania, USA, 16802}
\affil[2]{Department of Mechanical and Production Engineering, Aarhus University, 8000 Aarhuc C, Denmark}

\maketitle

\noindent\makebox[\linewidth]{\rule{\linewidth}{0.6pt}}

\begin{abstract}
Calibrating a Reynolds-averaged Navier-Stokes (RANS) model against data leads to an improvement.
Determining {\it a priori} if such an improvement generalizes to flows outside the calibration data is an outstanding challenge.
This work attempts to address this challenge via global epistemic Uncertainty Quantification (UQ).
Unlike the available epistemic UQ methods that are local and tell us a model's uncertainty at one specific flow condition, 
the global epistemic UQ method presented in this work tells us also whether a perturbation of the original model would generalize.
Specifically, the global epistemic UQ method evaluates a potential improvement in terms of its ``effectiveness'' and ``inconsistency''.
Any improvement can be put in one of the following four quadrants: first, high effectiveness, low inconsistency; second, high effectiveness, high inconsistency; third, low effectiveness, low inconsistency; and fourth, low effectiveness, high inconsistency.
An improvement would generalizes if and only if it is in the high effectiveness and low inconsistency quadrant.
To demonstrate the concept, we apply the global epistemic UQ to full Reynolds stress modeling of a stratified shear layer.
The global epistemic UQ results point to a model coefficient in the pressure-strain correlation closure (among others) as effective and consistent for predicting the quantity of interest of shear layer's growth.
We calibrate the model coefficient such that our RANS matches direct numerical simulation data at one flow condition.
We show that the calibrated model generalizes to several other test flow conditions.
On the other hand, when calibrating a high inconsistency term, we get a model that works at only the calibrated condition.
\end{abstract}

\noindent\makebox[\linewidth]{\rule{\linewidth}{0.6pt}}

\section{Introduction}
\label{sect:intro}

Predicting turbulence and its impact is critical to many applications in science and engineering.
For example, predicting stratified wake flow is required in numerous applications including atmospheric and marine meteorology, and undersea vehicle motion \cite{lin1979wakes}.
Limited by the available computational resources, direct numerical simulation (DNS) and large-eddy simulation (LES) of high Reynolds number and/or high Richardson number stratified wake flow are and will remain infeasible in the foreseeable future \cite{choi2012grid,yang2020grid}. 
Accordingly, Reynolds-averaged Navier-Stokes (RANS) modeling is the go-to tool for industrial-scale flow problems.

At a high level, a RANS model represents a mapping between the mean flow and the underlying turbulence.
Because there is not a universal mapping between the mean flow and the underlying turbulence \cite{frisch1995turbulence,spalart2015philosophies}, a specific RANS model and a specific set of model coefficients can exhibit suitable accuracy for only a limited range of applications.
As pointed out in Ref \cite{spalart2015philosophies}, it is hard to imagine that a RANS model calibrated to only isotropic turbulence would have predictive power over, e.g., complicated flows like stratified wakes.
Hence, it is a common practice to calibrate a RANS model to a training flow before it is used for a flow of interest or a test flow.
For instance, the full Reynolds stress model (FRSM) is calibrated to zero-pressure-gradient boundary-layer flow before applied to other wall-bounded flows \cite{mellor1974hierarchy}.
In a recent study, different RANS models are calibrated to a boundary layer with a moderate adverse pressure gradient before they are applied to a boundary layer with a high adverse pressure gradient \cite{volino2020non}. 

Calibrating a RANS model requires data.
At the time when the k-$\epsilon$ model, the k-$\omega$ model, and the Spalart-Allmaras (SA) model were developed \cite{jones1972prediction,menter1994two,spalart1992one}, there was not a lot of high-quality data that would be used for model calibration.
Things are, of course, different nowadays.
The past decade has seen much progress in high-performance computing \cite{messina2017exascale}, and this increased computational power has enabled DNS and LES of flows at moderate and moderately high Reynolds numbers \cite{graham2016web,lee2015direct, yamamoto2018numerical}. 
These DNS and LES data provide new opportunities for RANS calibration and RANS modeling in general.
A notable example is the use of machine learning in RANS models calibration \cite{duraisamy2015new,tracey2015machine,ling2016machine,ling2016reynolds,ray2016bayesian,wang2017physics,wu2018physics,zhang2019recent,milani2020turbulent,zhao2020rans}.
The general methodology is to train a baseline RANS model to high-fidelity data.
The baseline RANS model is usually retained, and machine learning improves upon that baseline RANS model.
There are many things one can do to make the training process more efficient.
For example, one can constrain the training so that the resulting RANS model is Galilean invariant and realizable \cite{ling2016machine,zhao2020rans}; one can also include as many inputs as possible to ensure that the resulting model does not underfit \cite{wu2018physics}.
This machine learning approach has seen many successes, and here, we name a few.
Tracey et al. \cite{tracey2015machine} showed that an artificial neural network recovers the SA model from data. 
Ling et al. \cite{ling2016reynolds} showed that embedding Galilean invariance in network training greatly reduces the amount of training data needed. 
Wu et al. \cite{wu2018physics} extended Ling et al.'s work and incorporated rotational invariance and reflectional invariance.
The main criticism of machine learning models, however, is the models' ability to generalize, or rather, a lack of knowledge about the model's applicable range.

A lack of knowledge of a model's applicable range is fundamentally due to a lack of understanding of the flow physics.
If one perfectly understands the flow physics, that understanding could well be used to determine whether a specific model, calibrated to a specific flow, generalizes to another specific flow.
Although having a perfect understanding of flow physics is nearly impossible, it is possible to understand and quantify RANS models' uncertainties.
That directly leads to epistemic uncertainty quantification (UQ).
The present literature categorizes epistemic uncertainties in terms of parametric model uncertainties and non-parametric model uncertainties.
Parametric uncertainties are the uncertainties in the model coefficients, and non-parametric uncertainties are the uncertainties in the model form.  
Both types of uncertainties can be studied by data-free methods and data-driven methods.
Specifically, data-free methods consist of propagating pre-specified probability distributions of model coefficients and model terms through RANS equations and investigating the uncertainty distribution of the solution, see Refs \cite{dunn2011uncertainty,margheri2014epistemic,schaefer2017uncertainty} for data-free UQ of parametric uncertainties and Refs \cite{emory2013modeling,iaccarino2017eigenspace,mishra2017uncertainty,edeling2018data,xiao2017random} for data-free UQ of non-parametric uncertainties. 
On the other hand, data-driven methods consist of assimilating data to infer the probability distributions of the model coefficients and model terms;
the inferred probability distributions are available for propagating through RANS equations in a subsequent prediction step; see Refs \cite{cheung2011bayesian,kato2013approach,edeling2014predictive,edeling2014bayesian,kato2015data,papadimitriou2015bayesian} for data-driven UQ of parametric uncertainties and Refs \cite{singh2016using,xiao2016quantifying,wu2016bayesian,wang2016incorporating,parish2016paradigm,gorle2019epistemic} for data-driven UQ of non-parametric uncertainties.
Although UQ of RANS is a somewhat new field, the research has led to many insights in the past few years.
{Here, we name a few.
Margheri et al. \cite{margheri2014epistemic} showed high-level uncertainty in $k-\epsilon$ and $k-\omega$ RANS models' channel results due to the uncertainties in the model coefficients.
Iaccarino et al. \cite{iaccarino2017eigenspace} obtained a bound for the $k-\omega$ model results of flow behind a back-facing step by perturbing the shape of the Reynolds stress (we will define the shape of the Reynolds stress in section \ref{subsec:physics}).
Cheung et al. \cite{cheung2011bayesian} found that their data-driven method leads to different model coefficients in the Spalart-Allmaras model than the conventional model coefficients.
Xiao et al. \cite{xiao2016quantifying} showed that more accurate estimates of fluid velocity can be obtained by assimilating information from data following an iterative ensemble Kalman method.
}
A more detailed review of the literature, e.g., Refs \cite{zadeh1996fuzzy,klir1996generalized,ferson2004arithmetic},  falls out of the scope of this work, and 
the reader is directed to Ref. \cite{xiao2019quantification}.
In all, the available epistemic UQ methods are local: they give uncertainty estimates at a given flow condition but provide very little information about a model's applicable range.

To more clearly see local UQ methods' limitations, let us consider RANS of a stratified shear layer.
A stratified shear layer is controlled by the Reynolds number, $Re$, and the Richardson number, $Ri$ (and other relevant parameters like the turbulence intensity $TI$).
These flow controlling parameters (FCPs) define a FCP space.
Let us say that we have a FRSM that does not accurately predict the shear layer's growth.
Let us also say that we have DNS data at one flow condition, i.e., one specific $Re$, $Ri$, and $TI$, and we want to make use of our DNS data to improve the RANS model.
A local UQ method would help if the objective is to improve the model's performance at the DNS's condition, but it falls short if the objective is to develop an improvement that generalizes to other flow conditions within the FCP space.

That being said, besides UQ, one can resort to {\it a posteriori} tests to determine if an improvement generalizes to a flow condition.
By comparing a model's prediction to data, {\it a posteriori} tests give us definite answer as whether an improvement generalizes to a specific flow condition or not \cite{rumsey2006summary,vassberg2007summary,eisfeld2016verification,roy2018summary}.
This process is also known as verification and validation.
The difficulty, however, is that verification and validation require high-fidelity data at the test conditions, which are often not available.

The objective here is to know {\it a priori} whether an improvement generalizes.
To know that, we need to know if the improvement has a consistent impact on a quantity of interest (QoI)---the improvement generalizes only if it has a consistent impact.
This leads directly to global epistemic UQ methods, i.e., the topic of this paper.
The rest of the paper is organized as follows.
The methodology is presented in section \ref{sect:method}.
In section \ref{sect:app}, we apply the global epistemic UQ method to FRSM of a stratified shear layer.
We conclude in section \ref{sect:conclusion}.

\section{Methodology}
\label{sect:method}

\subsection{Problem statement}
\label{subsec:prob}

By definition, no RANS model can fully account for the flow physics, and this leads to model uncertainty, the effects of which can be studied by perturbing the model terms. 
If one calibrates a model to data such that the calibrated model returns a more accurate prediction of a QoI than the original model, the calibrated model can be considered a model improvement.
The improvement may be regarded as a perturbation to the original model.
We want to know the improvement's applicable range.

The problem concerns two spaces: the FCP space and the model space.
A point in the FCP space defines a flow.
For example, a stratified shear layer is controlled by the Richardson number and the Reynolds number.
For a stratified shear layer, the Richardson number and the Reynolds number constitute a two-dimensional FCP space, and a point in this FCP space, i.e., a specific combination of the Richardson number and the Reynolds number, defines a stratified shear layer.
Similarly, a point in the model space defines a RANS model.
For example, a FRSM requires closures for the pressure-strain term, the dissipation term, the diffusion term, etc. (see details in section \ref{sub:FRSM}). 
These closure models and the constants in these models constitute the model space.
A point in this model space corresponds to a specific set of model closures and a specific set of model constants, which together define a FRSM model.
Given a QoI, a FCP space, and a model space, a global epistemic UQ method maps uncertainties in the model space to uncertainties in the QoI for flows in the FCP space.

\subsection{General methodology}
\label{subsec:method}

\begin{figure}
    \centering
    \includegraphics[width=1.0\textwidth]{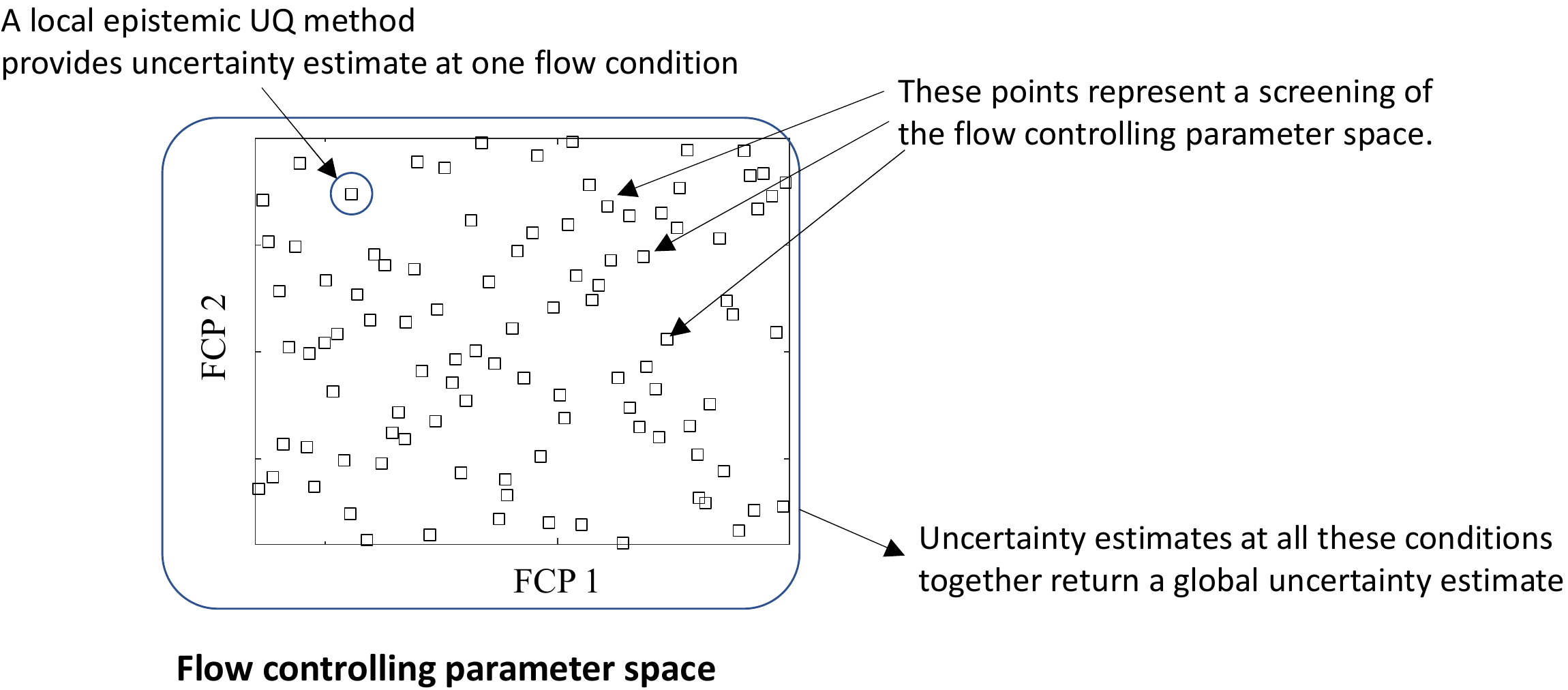}
    \caption{A sketch of the general methodology.
    The sampling points in the 2D space is generated according to the ``constrained'' Latin hypercube method (see section \ref{subsec:FCS} for details). }
    \label{fig:FCS-sketch}
\end{figure}

Figure \ref{fig:FCS-sketch} is a sketch of the general methodology.
The global epistemic UQ method consists of a local epistemic UQ method and a method that screens the FCP space.
A screening of the FCP space gives $N$ sampling points.
The local epistemic UQ method provides uncertainty estimates at these $N$ points. 
These scalar uncertainty values are defined as $e_i$, $i=1$, 2, 3, ..., $N$, for $N$ flow conditions.
These local uncertainty estimates together return a global uncertainty estimate.

\begin{figure}
    \centering
    \includegraphics[width=0.5\textwidth]{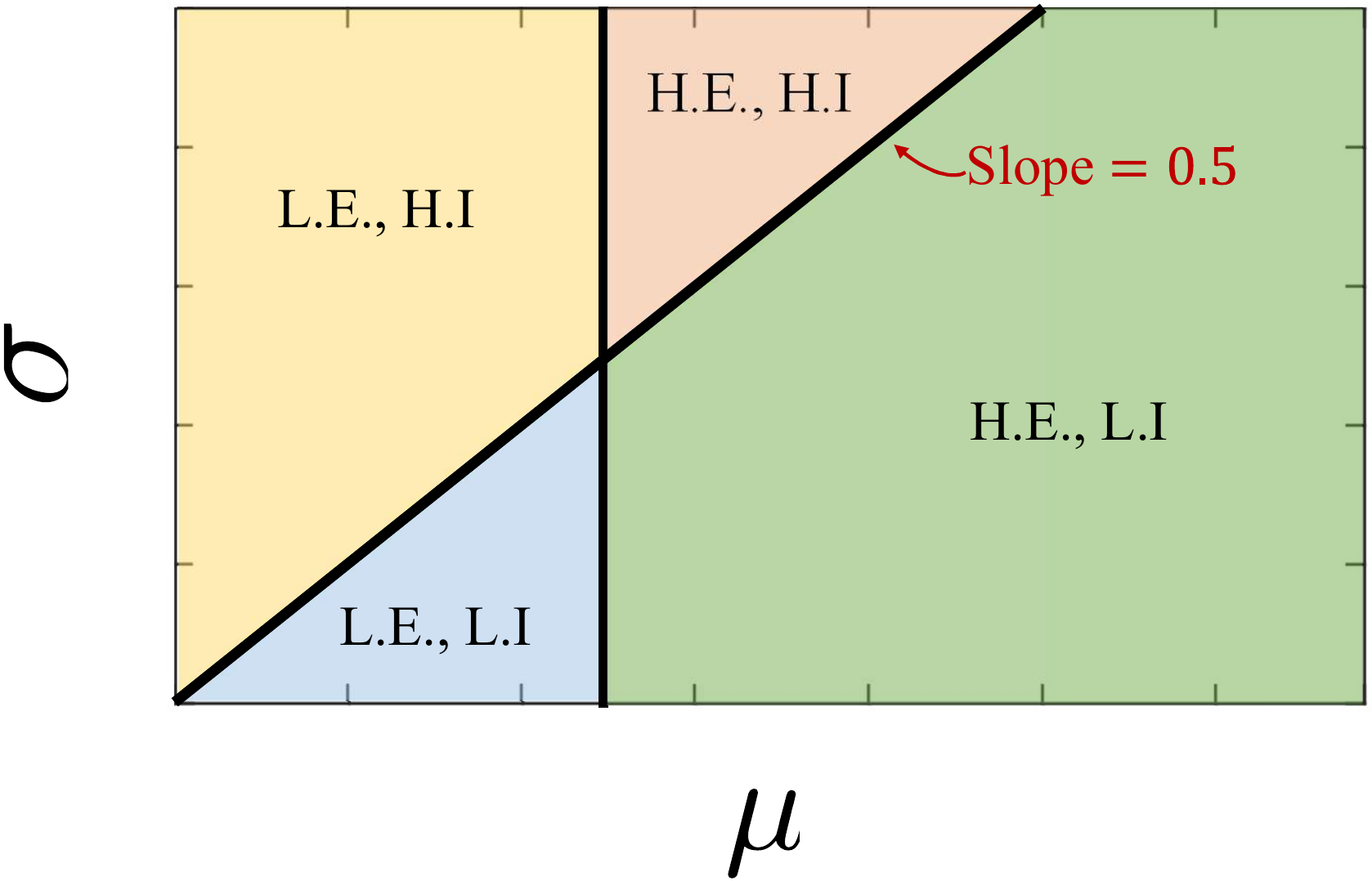}
    \caption{$\sigma-\mu$ quadrant diagram. 
    $\mu$ measures a perturbation's effectiveness.
    $\sigma$ measures a perturbation's inconsistency.
    Large $\mu$ value is high effectiveness, i.e., H.E.. 
    Small $\mu$ value is low effectiveness, i.e., L.E..
    Large $\sigma$ value is high inconsistency, i.e., H.I.. 
    Small $\mu$ value is low inconsistency, i.e., L.I..
    Here, if $\sigma/\mu>~$0.5, we say $\sigma$ has a large value, and if $\sigma/\mu<~$0.5, we say $\sigma$ has a small value.}
    \label{fig:sigma-mu-sketch}
\end{figure}

Following Ref \cite{emory2013modeling}, we study a model's uncertainty by perturbing the model terms.
We measure a perturbation in terms of its ``inconsistency'' and ``effectiveness''. 
Effectiveness is measured according to 
\begin{linenomath*}\begin{equation}
    \mu=\frac{1}{N}\sum_{i=1}^N e_i,
    \label{eq:def-mu}
\end{equation}\end{linenomath*}
i.e., the mean of $e_i$,
and inconsistency is measured according to
\begin{linenomath*}\begin{equation}
    \sigma^2=\frac{1}{N}\sum_{i=1}^N (e_i-\mu)^2,
    \label{eq:def-sigma}
\end{equation}\end{linenomath*}
i.e., the standard deviation of $e_i$.
Equations \eqref{eq:def-mu} and \eqref{eq:def-sigma} define a quadrant diagram, as sketched in figure \ref{fig:sigma-mu-sketch}.
If $\mu$ is large and $\sigma$ is small, $e_i$'s have large values that vary little in the FCP space.
This is high effectiveness (H.E.), low inconsistency (L.I.).
It is worth noting here that $\sigma$'s value must be compared to $\mu$: a $\sigma$'s value is considered large only if it is large compared to $\mu$, and a $\sigma$'s value is considered small only if it is small compared to $\mu$.
Similarly, we can define the H.E., high inconsistency (H.I.) quadrant, the low effectiveness (L.E.), L.I. quadrant, and the L.E., H.I. quadrant.
To further illustrate the concept, let us consider the following example.
Let us say that we have $i$=1, 2, i.e., the FCP space contains two sampling points.
Let us also say that $\mu=10$ is considered high effectiveness, $\mu=1$ is considered low effectiveness, $\sigma/\mu=0$ is considered low inconsistency, and $\sigma/\mu=1$ is considered high inconsistency.
If a type of perturbation leads to $e_1=1$, $e_2=1$, we have $\mu=1$, $\sigma/\mu=0$, and the perturbation is in the L.E., L.I. quadrant.
Because of its low effectiveness, such a perturbation is not likely to lead to any significant improvement.
The same is true if the perturbation is in the L.E., H.I. quadrant.
If a perturbations leads to $e_1=5$, $e_2=15$, we have $\mu=10$, $\sigma/\mu=1$, and the perturbation is in the H.E., H.I. quadrant.
This perturbation will lead to improvement at the condition $i=2$, but because of its high inconsistency, this perturbation would not generalize and work well at the condition $i=1$.
Now, if a perturbation leads to $e_1=10$, $e_2=10$, we have $\mu=10$, $\sigma/\mu=0$, and the perturbation is in the H.E., L.I. quadrant.
We will show in section \ref{sub:app-results} that an improvement will generalize only if it is in the H.E., L.I. quadrant.

\subsection{Cost of global epistemic UQ}
\label{subsec:cost}

Having presented the general methodology, we now estimate the cost of global epistemic UQ.
The cost of the global epistemic UQ method is roughly proportional to the times we need to evaluate the RANS model.
If $N'$ evaluations (i.e., RANS runs) are needed for local UQ at one flow condition, global epistemic UQ requires $NN'$ times' evaluation of the RANS model.
Again, $N$ is the number of samplings in the FCP space, e.g., flow conditions.
In the following, we estimate $N$ and $N'$.
 
Let us consider a $d$-dimensional FCP space and a $d'$-dimensional model space.
The value of $d$ typically ranges from O(1) to O(10).
For a stratified shear layer, the Reynolds number and the Richardson number gives a $d=2$ dimensional FCP space.
The dimension of the FCP space will be higher if geometric parameters are involved.
For example, for a stratified wake, geometric parameters that control the geometry of the wake-generating body add to FCP's dimension.
The value of $d'$ typically ranges from O(10) to O(100) depending on the number of unclosed terms and the number of model coefficients.
For an eddy viscosity model, $d'$ is about 10 \cite{spalart1992one}. 
$d'$ is large for more complicated models like FRSM.
For a FRSM, e.g., the one detailed in section \ref{sect:app}, $d'$ is approximately 100.

If one employs a space-filling strategy \cite{santner2003design}, screening the FCP space and the model space requires $N\sim n^d$ and $N'\sim n'^{d'}$ samples, where $n$ and $n'$ are O(10) constants, which is prohibitively costly.
To elaborate, let us consider a 2D FCP space and a (tiny) 6D model space.
Taking $n=10$ and $n'=10$ already requires $10^8$ times' evaluation of the RANS model.
If more efficient sampling strategies are employed, e.g., the ones described in sections \ref{subsec:modelspace} and \ref{subsec:FCS}, screening the two space requires $N\sim n$ and $N'\sim n'd'$ samples respectively.
In that case, global epistemic UQ requires O(100) times' evaluation of the RANS model for a 2D FCP space and a 6D model space.

\subsection{Screening model space}
\label{subsec:modelspace}

\begin{figure}
    \centering
    \includegraphics[width=0.5\textwidth]{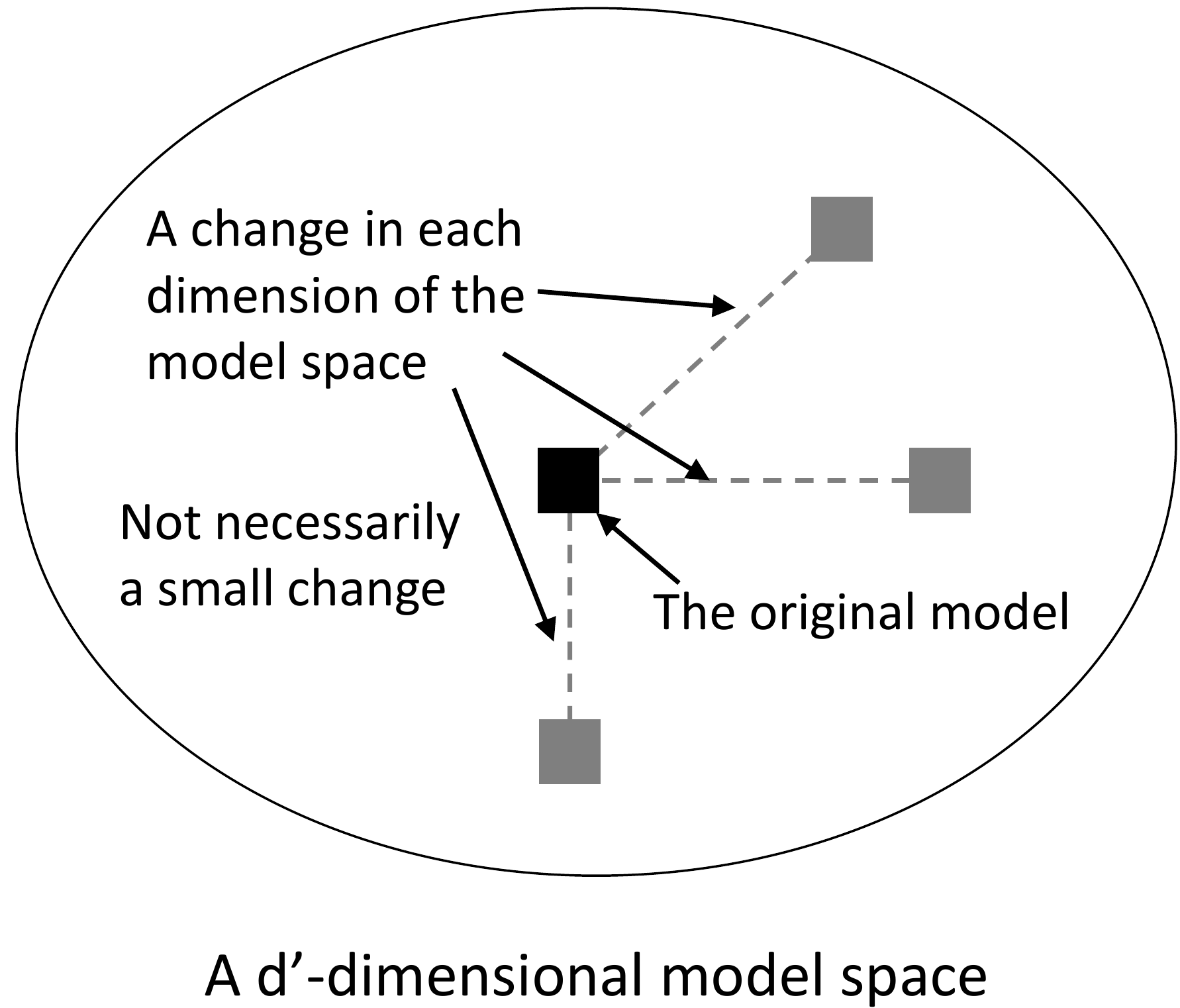}
    \caption{A sketch of the local UQ method.}
    \label{fig:localUQ-sketch}
\end{figure}

We perturb the original RANS model such that the perturbed models lie in the vicinity of, but not necessarily close to the original model in the model space.
Figure \ref{fig:localUQ-sketch} is a sketch of the method.
Here, except for the original model, a point in the model space differs from the original model in only one dimension.
This method requires $d'+1$ evaluations of the RANS model for local UQ.

Compared to the information we can get from the recently-developed more-advanced local epistemic UQ methods like the ones in Ref \cite{edeling2014predictive}, we get less information from the method described here.
Although, in principle, we can use the recently-developed methods for local UQ, these methods are more costly.
If we were to follow, e.g., Ref \cite{edeling2014predictive}, and propagate the PDFs of all model coefficients and all model terms through a RANS model, our global epistemic UQ would be O(100) times more costly.
Considering that our methodology already requires about O(1000) RANS calculations, employing the recently-developed more-advanced local UQ methods like the one in Ref \cite{edeling2014predictive} is not practical.

\subsection{Screening flow controlling parameter space}
\label{subsec:FCS}

\begin{figure}
    \centering
    \includegraphics[width=0.63\textwidth]{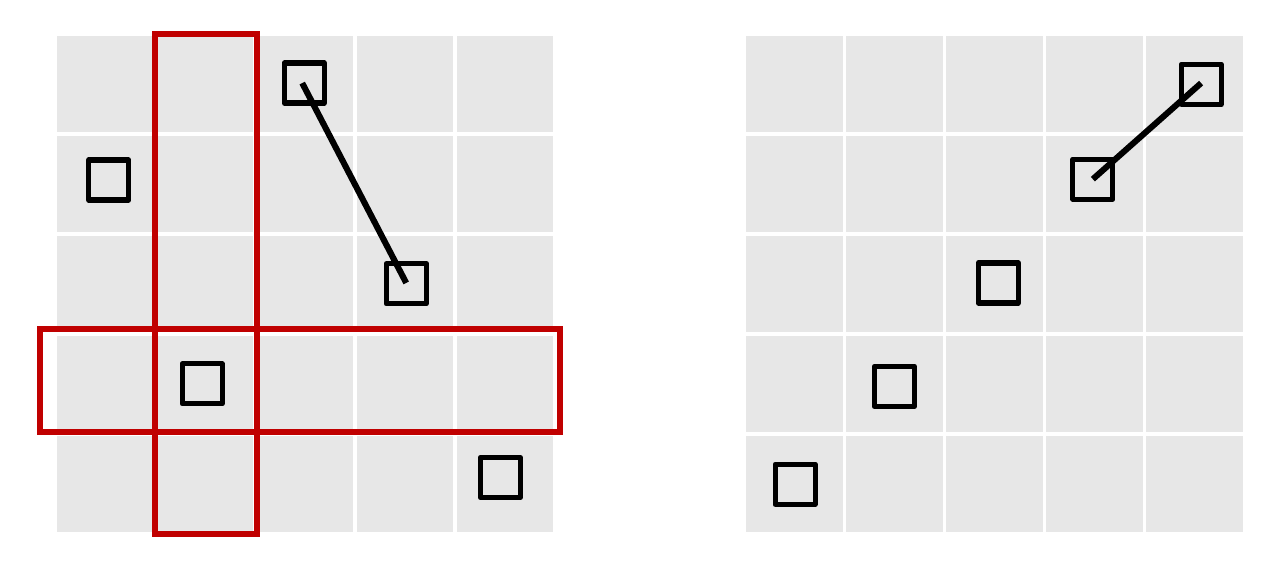}
    \caption{Two possible Latin hypercube designs.
    The space is two dimensional (for visualization purposes).
    The thin white lines define a square grid.
    The black squares are sampling points.
    The red boxes are two hyperplanes.
    For a 2D space, a hyperplane is a 1D line.
    The two black lines indicate the smallest inter-point distance in the two designs.
    }
    \label{fig:LH-sketch}
\end{figure}

To keep the cost manageable, we employ the ``constrained'' Latin hypercube strategy to screen the FCP space. 
Details of the method can be found in Ref. \cite{morris1995exploratory}.
Here, we briefly summarize the main features of the method.
First, we define a Latin hypercube.
A $d$-dimensional square grid containing sample positions is a Latin hypercube if and only if each sample is the only one in axis-aligned hyperplanes containing it.
When sampling a $d$-dimensional space, the range of each dimension is divided into $N$ equally probable intervals. 
$N$ sample points are then placed to satisfy the Latin hypercube requirements.
Figure \ref{fig:LH-sketch} shows two possible ``designs'' in a 2D space.
Here, a design is a realization of the Latin hypercube.
Although both designs are Latin hypercubes, the design in figure \ref{fig:LH-sketch} (a) is a more explorative one than the one in figure \ref{fig:LH-sketch} (b), making the design in figure \ref{fig:LH-sketch} (a) a better design.
This motivated Morris and Mitchell \cite{morris1995exploratory} to constrain the Latin hypercube such that the distances between any two samples are maximized, leading to the ``constrained'' Latin hypercube method.

\subsection{Physics-informed perturbations of the original model}
\label{subsec:physics}

Any change/perturbation of the original model in the model space should respect the basic physics.
For example, it is desirable that if one perturbs a symmetric (or/and positive definite) tensor, the perturbed tensor remains symmetric (or/and positive definite) \cite{wu2018physics}. 
Following Ref \cite{emory2013modeling}, perturbations of a tensor are defined in terms of the tensor's magnitude, as well as shape and orientation of the normalized anisotropic part of the tensor:
\begin{linenomath*}\begin{equation}
    a_{ij}=a_{kk}\left(\frac{1}{3}\delta_{ij}+V_{ik}\Lambda_{kl}V_{lj}\right),
\end{equation}\end{linenomath*}
where $a$ is an arbitrary symmetric (positive definite) tensor, $V$ is the eigenvector matrix, and $\Lambda$ is the eigenvalue matrix.
Perturbing a tensor's magnitude is to vary its trace $a_{kk}$, which is one dimensional;
perturbing a tensor's shape is to vary its eigenvalues $\Lambda$, which is two dimensional;
and perturbing a tensor's orientation is to vary a tensor's eigenvectors $V$, which is three dimensional.
Hence, perturbing a symmetric (positive definite) tensor gives rise to a 6 dimensional model space.
In addition to perturbations to a tensor term, one can also define perturbations to a vector term in terms of the vector's magnitude and its orientation:
\begin{linenomath*}\begin{equation}
    b_i=\sqrt{b_kb_k}\hat{b}_i,
\end{equation}\end{linenomath*}
where $b_i$ is an arbitrary vector, $\hat{b}_i$ is a unit vector in the direction of $b_i$.
Perturbing a vector's magnitude is to vary its length $\sqrt{b_kb_k}$, which is one dimensional;
perturbing a vector's direction is to vary $\hat{b}_i$, which is two dimensional.
Hence, perturbing a vector gives rise to a 3 dimensional model space.
Thus defined perturbations to tensors and vectors are Galilean invariant.
Perturbations of a scalar term are always Galilean invariant.

\subsection{Procedure of global epistemic uncertainty quantification}

The analysis begins with a baseline RANS model, a model space, a FCP space, and a QoI.
We define $e_{i,j,{\rm QoI}}$: a specific type of perturbation $j$ to the original RANS model at the flow condition $i$ gives rise to an uncertainty estimate $e_{i,j,{\rm QoI}}$ for the QoI.
Here, $i=1$,  2, 3, ..., $N$, and $j=$1, 2, 3, ..., $d'$.
The steps of a global epistemic UQ are as follows.
\begin{itemize}
    \vspace{-2mm}
    \item[1.] sample the FCP space $N$ times following the method in section \ref{subsec:FCS}.
    \vspace{-2mm}
    \item[2.] compute $e_{i,j,{\rm QoI}}$ for $i$ from $1$ to $N$ for a fixed $j$.
    \vspace{-2mm}
    \item[3.] compute $\mu_j$ and $\sigma_j$ according to 
    \begin{linenomath*}\begin{equation}
    \mu_j=\frac{1}{N}\sum_{i=1}^N e_{i,j,{\rm QoI}},~~~~
    \sigma_j^2=\frac{1}{N}\sum_{i=1}^N (e_{i,j,{\rm QoI}}-\mu_j)^2.
    \label{eq:sigma-mu}
    \end{equation}\end{linenomath*}
    \vspace{-2mm}
    \item[4.] repeat the above three steps for $j=1$, 2, 3, ..., $d'$.
    \vspace{-2mm}
\end{itemize}
The effectiveness and inconsistency, i.e., $\mu_j$ and $\sigma_j$, contain the global uncertainty information. 
A specific type of perturbation $j$ to the original model generalizes in the FCP space if and only if it is in the H.E., L.I. quadrant, i.e., only if it has a large $\mu$ value and small $\sigma/\mu$ value. 

\subsection{Discussion}
\label{subsect:discussion}

In this subsection, we remark on the global epistemic UQ method.

\vspace{2mm}
\noindent
{\bf Remark 1} We are not introducing a new sampling method.
The local epistemic UQ method in section \ref{subsec:modelspace} and the global screening method in section \ref{subsec:FCS} are well-established methods and require no further validation. 
We combine established methods to address an open challenge.

\vspace{2mm}
\noindent
{\bf Remark 2} Because of the use of the $\sigma-\mu$ diagram and local epistemic method in section \ref{subsec:modelspace}, an informed reader would find the global epistemic UQ method reminiscent of the Morris method.
Although the difference between the two methods will become clear in section \ref{sect:app}, it is worth noting that our global epistemic UQ method differs from the Morris method in terms of the working space.
The Morris method involves one working space, whereas our global epistemic UQ method involves two, i.e., the FCP space and the model space.

\vspace{2mm}
\noindent
{\bf Remark 3} In the current CFD literature, epistemic UQ's overarching objective is to give a bound for RANS results.
Consequently, most work does not separately study the effects of different types of model perturbations.
This is not an issue for eddy viscosity models, because all one can perturb is the Reynolds stress term.
However, when many unclosed terms represent different physical processes in a complicated model like FRSMs, not separately studying each perturbation's effect leads to a significant loss of information.
Here, we prevent this loss of information by computing $\sigma_{j}$ and $\mu_{j}$ for $j=1$, 2, ... $d'$ separately.

\section{Application}
\label{sect:app}

We begin our discussion by stating the general problem.
We have DNS data of a temporally-evolving shear layer at one flow condition and a RANS model that does poorly in terms of predicting the shear layer's growth.
We want to calibrate the RANS model such that the calibrated model more accurately predicts the shear layer's growth at the DNS's condition and other conditions within a designated FCP space.

The rest of the section is organized as follows.
We review the flow phenomenology of a temporally-evolving stratified shear layer in section \ref{sub:app-ssl}.
The details of the FSRM are presented in section \ref{sub:FRSM}.
The details of the UQ method specific to this application are presented in section \ref{sub:app-details} followed by the results in section \ref{sub:app-results}.
In section \ref{sub:app-ml}, we calibrate the FRSM and show that the calibrated model generalizes.
The reader can skip the first two subsections if he/she is familiar with the RANS model and the flow problem.
Details of the RANS and DNS codes and the simulation setup are presented in \ref{app:RANS} and \ref{app:DNS}.

\subsection{Stratified shear layer}
\label{sub:app-ssl}

\begin{figure}
    \centering
    \includegraphics[width=1.0\textwidth]{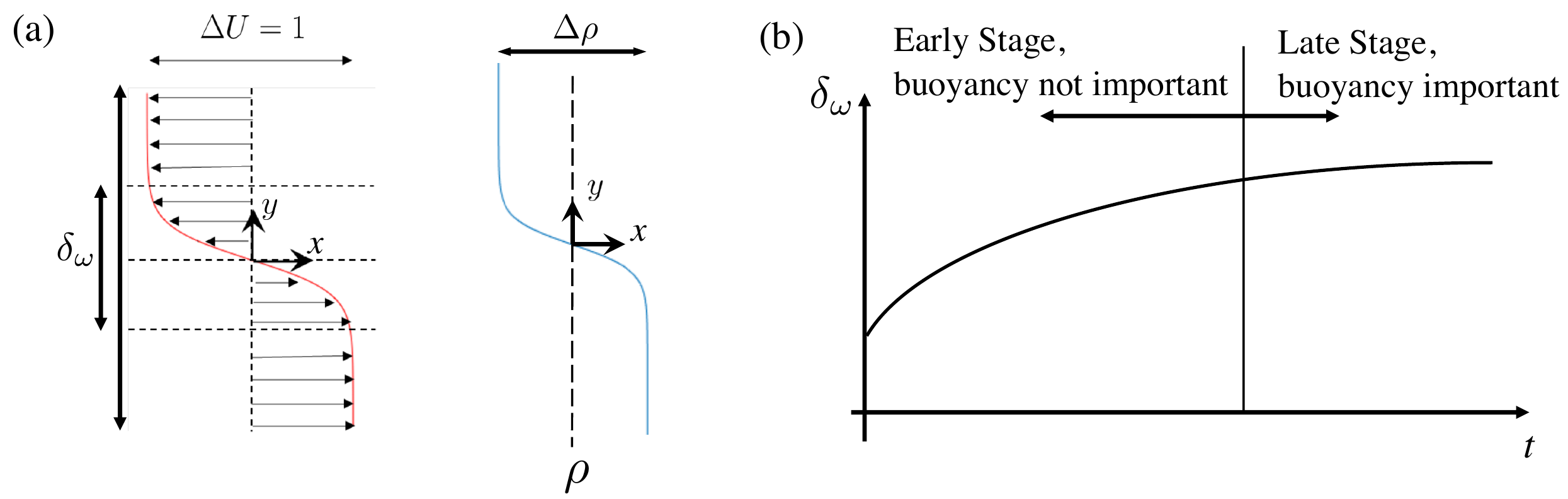}
    \caption{(a) A sketch of a stably stratified shear layer. 
    A stably stratified shear layer emerges between two streams moving at the same speed but in two opposite directions.
    The fluid velocity is in the $x$ direction.
    The velocity difference between the two streams is $\Delta U=1$ (unit).
    The top stream is lighter than the bottom stream.
    The density difference between the two streams is $\Delta \rho$.
    (b) A sketch of the growth of a stratified shear layer.
    At the early stage, buoyancy is not important.
    At the late stage, buoyancy becomes important.
    }
    \label{fig:shear}
\end{figure}

Figure \ref{fig:shear} (a) is a sketch of a temporally-evolving stably stratified shear layer.
A temporally-evolving stratified shear layer emerges between two streams of flows that  move at the same speed but in the opposite directions, with the top stream lighter than the bottom stream.
Temporally evolving stratified shear layer flow has been a model problem for flows with thermal stratification, and its phenomenology has received much attention in the published literature \cite{basak2006dynamics,brucker2007evolution,pham2009dynamics,pham2010transport,pham2014large}.
The initial state of the flow is controlled by the Reynolds number $Re$ \begin{linenomath*}\begin{equation}
    Re=\frac{\Delta U\delta_{\omega,0}}{\nu},
\end{equation}\end{linenomath*}
and the Richardson number $Ri$, 
\begin{linenomath*}\begin{equation}
    Ri\equiv \frac{g\Delta \rho\delta_{\omega,0}}{\rho \Delta U^2},
\end{equation}\end{linenomath*}
where $\Delta U$ is the velocity difference between the two streams, $\delta_{\omega,0}=\Delta U/\max[dU/dy]$ chracterizes the shear layer thickness, $g$ is gravity, $\Delta \rho$ is the density difference between the two streams, and $\rho$ is the reference density. 
Figure \ref{fig:shear} (b) is a qualitative sketch of stably stratified shear layer growth.
When the shear layer is thin, the flow is at the ``early stage'', where buoyancy does not play a very important role.
As the shear layer grows, the flow enters the ``late stage'', and buoyancy significantly attenuates the shear layer's growth.

\subsection{Full Reynolds stress model}
\label{sub:FRSM}

Linear eddy viscosity RANS models perform poorly for stratified shear layer flow because of the dynamically important Reynolds stress anisotropy that arises in these flows \cite{launder1989second}.
FRSMs solve the Reynolds stress transport equations, and thereby inherently accommodate Reynolds stress anisotropy. 
In these models, the dissipative, diffusive, and redistributive processes in the Reynolds stress transport equations are modeled \cite{durbin2018some}. 
{In systems where buoyancy plays an important role, the model also incorporates transport equations for the turbulent temperature fluxes, i.e., $\left<u'_i\theta'\right>$ and temperature variance, i.e., $\left<\theta' \theta'\right>$.}
A significant literature exists for these relatively complex models, see e.g., \cite{shir1973preliminary,gibson1978ground,craft1996recent,eisfeld2016verification,eisfeld2019steps,eisfeld2020length}. 
Here we provide an overview of this modeling approach.
The RANS equations with the Boussinesq approximation can be written as
\begin{linenomath*}\begin{equation}
\begin{split}
    \frac{\partial U_i}{\partial x_i} &= 0,\\
    \frac{\partial U_i}{\partial t} + U_j \frac{\partial U_i}{\partial x_j} &= -\frac{1}{{\rho}}\frac{\partial P}{\partial x_i} + \nu\frac{\partial^2 U_i}{\partial x_j \partial x_j} - \frac{\partial\langle{u_i'u_j'}\rangle}{\partial x_j} - \beta g_i \Theta, \\
    \frac{\partial \Theta}{\partial t} + U_j \frac{\partial \Theta}{\partial x_j} &= \alpha\frac{\partial^2 \Theta}{\partial x_j \partial x_j} - \frac{\partial\langle{\theta'u_j'}\rangle}{\partial x_j},
\end{split}
\label{eq:RANS_mean}
\end{equation}\end{linenomath*}
where $U_i$ is the Reynolds-averaged velocity, $x_i$ is the coordinate, $t$ is time, $P$ is the Reynolds-averaged pressure, $\beta$ is the thermal expansion coefficient {at constant pressure}, $\Theta$ is the Reynolds-averaged temperature, $\alpha$ is the {fluid conductivity}, $\theta'$ is the temperature fluctuation, $u'_i$ is the velocity fluctuation, and $\left<\right>$ denotes a Reynolds average. The exact transport equations for the Reynolds stresses, $R_{ij}=\left< u'_i u'_j \right>$, temperature fluxes, $R_{\theta i}=\left< u'_i \theta' \right>$, and temperature variance, $R_{\theta \theta}=\left< \theta' \theta' \right>$, are as follows
\begin{linenomath*}\begin{align}\begin{split}
    \frac{\partial { R_{ij}}}{\partial t} + U_k\frac{\partial { R_{ij}}}{\partial x_k} & =  \underbrace{\left(-R_{ik} \frac{\partial U_j}{\partial x_k} -R_{jk} \frac{\partial U_i}{\partial x_k}\right)}_{\mathcal{P}_{i j}} + \underbrace{\left( \frac{\partial}{\partial x_k} \left( -\langle u'_i u'_j u'_k \rangle - \frac{1}{\rho}\left\langle p'\left(u'_i \delta_{jk} + u'_j \delta_{ik}\right) \right\rangle + \nu \frac{\partial R_{ij}}{\partial x_k} \right) \right)}_{\mathcal{D}_{i j}}\\ 
    & + \underbrace{\left( -\beta \left(g_i R_{\theta j} + g_j R_{\theta i}  \right) \right)}_{\mathcal{G}_{i j}} + \underbrace{\left( \left\langle \frac{p'}{\rho}\left( \frac{\partial u'_i}{\partial x_j} + \frac{\partial u'_j}{\partial x_i}\right) \right\rangle \right)}_{\Phi_{i j}} - \underbrace{2\nu\left\langle \frac{\partial u_i'}{\partial x_k}\frac{\partial u_j'}{\partial x_k} \right\rangle}_{\epsilon_{ij}},
    \label{eq:Rij}
\end{split}\end{align}\end{linenomath*}
\begin{linenomath}\begin{align}\begin{split}
        \frac{\partial R_{\theta i}}{\partial t} + U_k\left( \frac{\partial R_{\theta i}}{\partial x_k} \right) & = \underbrace{\left( -R_{ki} \frac{\partial \Theta }{\partial x_k} - R_{\theta k} \frac{\partial U_i}{\partial x_k}\right)}_{\mathcal{P}_{\theta i}} \\
       & + \underbrace{\left( \frac{\partial}{\partial x_k} \left( -\langle \theta' u'_i u'_k \rangle -\frac{1}{\rho} \langle p'\theta'  \rangle \delta_{ik} + \nu \left \langle \theta'\frac{\partial u'_i}{\partial x_k} \right\rangle + \alpha \left \langle u'_i\frac{\partial \theta'}{\partial x_k}\right\rangle\right) \right)}_{\mathcal{D}_{\theta i}}\\
       & + \underbrace{\left(- \beta g_i R_{\theta \theta} \right)}_{\mathcal{G}_{\theta i}} + \underbrace{\left( \frac{1}{\rho} \left \langle p' \frac{\partial \theta'}{\partial x_i} \right \rangle \right)}_{\Phi_{\theta i}} - \underbrace{(\nu + \alpha)\left\langle \frac{\partial \theta'}{\partial x_k}\frac{\partial u_i'}{\partial x_k} \right\rangle}_{\epsilon_{\theta i}}
       \label{eq:R_theta_i},
    \end{split}\end{align}\end{linenomath}
\begin{equation}
        \frac{\partial R_{\theta \theta}}{\partial t} + U_k\left( \frac{\partial R_{\theta \theta}}{\partial x_k} \right) = \underbrace{\left( -2R_{\theta k}\frac{\partial \Theta}{\partial x_k} \right)}_{\mathcal{P}_{\theta \theta}}+ \underbrace{\left( \frac{\partial}{\partial x_k} \left( \alpha \frac{\partial R_{\theta \theta}}{\partial x_k} - \left \langle u'_k \theta' \theta' \right \rangle \right) \right)}_{\mathcal{D}_{\theta \theta}} - \underbrace{2\alpha \left \langle \frac{\partial \theta'}{\partial x_k}\frac{\partial \theta'}{\partial x_k} \right\rangle}_{\epsilon_{\theta \theta}}
        \label{eq:R_theta_theta},
    \end{equation}
where $\mathcal{P}_{ij}$, $\mathcal{D}_{ij}$, $\mathcal{G}_{ij}$ are the Reynolds stress production, diffusion, and buoyant production tensors respectively, $\Phi_{ij}$ is the pressure-redistribution tensor, and $\epsilon_{ij}$ is the turbulence dissipation rate tensor. 
$\mathcal{P}_{\theta i}$, $\mathcal{D}_{\theta i}$, $\epsilon_{\theta i}$, $\mathcal{G}_{\theta i}$ are the temperature flux production, diffusion, dissipation, and buoyant production vectors respectively, and $\Phi_{\theta i}$ is the pressure-scrambling vector. 
$\mathcal{P}_{\theta \theta}$, $\mathcal{D}_{\theta \theta}$, $\epsilon_{\theta \theta}$ are the temperature variance production, diffusion and dissipation terms respectively. 
The production terms and the buoyancy terms, $\mathcal{P}_{ij}, \mathcal{P}_{\theta i}, \mathcal{P}_{\theta \theta}, \mathcal{G}_{ij}, \mathcal{G}_{\theta i}$, do not need to be modeled at the second moment closure level, as they are comprised of terms that are explicitly solved for in the model. 
The remaining terms on the right-hand side of Eqs. \eqref{eq:Rij}, \eqref{eq:R_theta_i}, \eqref{eq:R_theta_theta} need to be modeled, as summarized below.

\subsubsection{Diffusion Terms Modeling}

The Reynolds stress diffusion term, $\mathcal{D}_{ij}$, is comprised of
\begin{linenomath*}\begin{equation}
\small
    \mathcal{D}_{ij} = \underbrace{\left(- \frac{\partial \left<u_i'u_j'u_k'\right>}{\partial x_k}\right)}_{\mathcal{D}_{T,ij}} + \underbrace{\left(- \frac{1}{\rho} \frac{\partial}{\partial x_k} \left( \left<p'u_j'\right>\delta_{ik} + \left<p'u_i'\right>\delta_{jk} \right)\right)}_{\mathcal{D}_{p,ij}} + \underbrace{\left(\nu \frac{\partial^2 \left\langle u'_i u'_j \right\rangle }{\partial x_k\partial x_k}\right)}_{ \mathcal{D}_{\nu,ij}},
\end{equation} \end{linenomath*} 
where $\mathcal{D}_{T,ij}$ represents turbulent diffusion, $\mathcal{D}_{p,ij}$ is the pressure-diffusion term, and $\mathcal{D}_{\nu,ij}$ represents viscous diffusion. 
We follow the standard approach and model the diffusion terms $\mathcal{D}_{p,ij}$ and $\mathcal{D}_{T,ij}$ together \cite{shir1973preliminary}
\begin{linenomath*}\begin{equation}
    \mathcal{D}_{p,ij}+\mathcal{D}_{T,ij}
    =\frac{\partial}{\partial x_k}\left[D\frac{k^2}{\varepsilon}\frac{\partial R_{ij}}{\partial x_k}\right],
    \label{eq:Dij}
\end{equation}\end{linenomath*}
where $D$ is a diffusion model coefficient.
Similarly, for temperature flux diffusion, $\mathcal{D}_{\theta i}$, we have \begin{linenomath*}\begin{equation}
    \mathcal{D}_{\theta i} = \underbrace{\left(- \frac{\partial \left<\theta'u_i'u_k'\right>}{\partial x_k}\right)}_{\mathcal{D}_{T,\theta i}} + \underbrace{\left(- \frac{1}{\rho} \frac{\partial \left( \left<p'\theta'\right>\delta_{ik} \right)}{\partial x_k} \right)}_{\mathcal{D}_{p,\theta i}} + \underbrace{\left(\nu \left \langle \theta'\frac{\partial u'_i}{\partial x_k} \right\rangle + \alpha \left \langle u'_i\frac{\partial \theta'}{\partial x_k}\right\rangle \right)}_{ \mathcal{D}_{\nu,\theta i}}.
\end{equation} \end{linenomath*}
Consistent with Eq. \eqref{eq:Dij}, we adopt
\begin{linenomath*}\begin{equation}
    \mathcal{D}_{\theta i} =\mathcal{D}_{p,\theta i}+\mathcal{D}_{\nu,\theta i}+\mathcal{D}_{T,\theta i}=\frac{\partial}{\partial x_k}\left[\left(\nu+\frac{1}{Pr_{tur,\theta i}}\frac{k}{\omega}\right)\frac{\partial R_{\theta i}}{\partial x_k}\right]
    \label{eq:Dti}
\end{equation}\end{linenomath*}
where $Pr_{tur,\theta i}$ is a turbulent Prandtl number.
For variance diffusion, $\mathcal{D}_{\theta \theta}$, {we adopt}
\begin{linenomath*}\begin{equation}
    \mathcal{D}_{\theta \theta} = \mathcal{D}_{\nu,\theta \theta}+\mathcal{D}_{T,\theta \theta}=\frac{\partial}{\partial x_k}\left[\left(\alpha+\frac{1}{Pr_{tur,\theta\theta}}\frac{k}{\omega}\right)\frac{\partial R_{\theta \theta}}{\partial x_k}\right]
    \label{eq:Dtt}.
\end{equation}\end{linenomath*}

\subsubsection{Pressure-Redistribution Modeling}

For the pressure-redistribution term, $\Phi_{ij}$, we use the Speziale-Sarkar-Gatski (SSG) {model} \cite{speziale_sarkar_gatski_1991} 
\begin{linenomath*}\begin{equation}
\begin{split}
\small
    \Phi_{ij}&=-\left(C_1\varepsilon+\frac{1}{2}C_{1}^*\mathcal{P}_{kk}\right)a_{ij} + C_2\varepsilon\left(a_{ik}a_{kj}-\frac{1}{3}a_{kl}a_{kl}\delta_{ij}\right) + \left(C_3-C_3^*\sqrt{a_{kl}a_{kl}}\right)kS_{ij}^* \\ 
    & + C_4k\left(a_{ik}S_{jk}+a_{jk}S_{ik}-\frac{2}{3}a_{kl}S_{kl}\delta_{ij}\right) + C_5k(a_{ik}W_{jk}+a_{jk}W_{ik})-C_{bf}\left(\mathcal{G}_{ij}-\frac{1}{3}\mathcal{G}_{kk}\delta_{ij}\right).
\end{split}
\label{eq:Phiij}
\end{equation}\end{linenomath*}
Here
\begin{linenomath*}\begin{equation*}
\small
    a_{ij} \equiv \frac{R_{ij}}{k}-\frac{2}{3}\delta_{ij},\ 
    S_{ij} \equiv \frac{1}{2}\left(\frac{\partial U_i}{\partial x_j}+\frac{\partial U_j}{\partial x_i}\right),\ 
    S_{ij}^* \equiv S_{ij}-\frac{1}{3}S_{kk}\delta_{ij},\ 
    W_{ij} \equiv \frac{1}{2}\left(\frac{\partial U_i}{\partial x_j}-\frac{\partial U_j}{\partial x_i}\right),
\end{equation*}\end{linenomath*}
and $C_1$, $C_1^*$, $C_2$, $C_3$, $C_3^*$, $C_4$, $C_5$, $C_{bf}$ are model coefficients.
In anticipation of the results presented in section \ref{sub:app-results}, we summarize the physics associated with each model term in Eq. \eqref{eq:Phiij}. Pressure-redistribution is conventionally decomposed into rapid, slow, and harmonic components, and in buoyant flows (as here), an isotropization of body forcing term \cite{pope2001turbulent}. The harmonic term is important only near a solid boundary and therefore is not of concern for a stratified shear layer.
In Eq. \eqref{eq:Phiij}, the first two terms model the slow pressure-redistribution (modeled from the return to isotropy hypothesis \cite{rotta1951statistische,lumley1977return}), the third to the fifth terms model rapid pressure-redistribution, and the last term models the isotropization of the buoyancy force \cite{launder1989prediction}.

\subsubsection{Pressure-Scrambling Modeling}

The pressure-scrambling term in the flux transport equations, $\Phi_{\theta i}$, is also decomposed into three components, $\Phi_{\theta i, 1}$, $\Phi_{\theta i, 2}$ and $\Phi_{\theta i, 3}$. We adopt a simple return-to-isotropy model \cite{monin1965symmetry} for $\Phi_{\theta i, 1}$, a basic isotropization of production (IP) model \cite{launder1989prediction} for $\Phi_{\theta i, 2}$, and quasi-isotropic model \cite{launder1975heat} for $\Phi_{\theta i, 3}$
\begin{linenomath*}\begin{equation}
    \Phi_{\theta i}=\underbrace{-C_{\theta 1}\frac{\varepsilon}{k}R_{\theta i}}_{\Phi_{\theta i, 1}} + \underbrace{C_{\theta 2}R_{\theta k}\frac{\partial U_i}{\partial x_k}}_{\Phi_{\theta i, 2}} + \underbrace{C_{\theta 3}\beta g_i R_{\theta\theta}}_{\Phi_{\theta i, 3}},
    \label{eq:Phiti}
\end{equation}\end{linenomath*}
where $C_{\theta 1}$, $C_{\theta 2}$, and $C_{\theta 3}$ are model coefficients.

\subsubsection{Dissipation Modeling}

For the Reynolds stress dissipation, we assume \textit{local isotropy}
\begin{linenomath*}\begin{equation}
    \varepsilon_{ij}=\frac{2}{3}\varepsilon\delta_{ij}, 
    \label{eq:epsij}
\end{equation}\end{linenomath*}
where $\varepsilon$ is the turbulent dissipation rate, which is related to the specific dissipation rate $\omega$ as
\begin{linenomath*}\begin{equation}
    \omega = \frac{\varepsilon}{C_\mu k},
\end{equation}\end{linenomath*}
where $C_\mu$ is a model coefficient (Prandtl-Kolmogorov constant). We solve the following transport equation for $\omega$ \cite{menter1994two}
\begin{linenomath*}\begin{equation}
    \frac{\partial \omega}{\partial t}+U_j\frac{\partial \omega}{\partial x_j}=\frac{\alpha_\omega \omega}{2k}\left(\mathcal{P}_{jj} + \mathcal{G}_{jj} \right)-\beta_\omega\omega^2+\frac{\partial}{\partial x_j}\left[\left(\nu+\sigma_\omega\frac{k}{\omega}\right)\frac{\partial \omega}{\partial x_j}\right]+\sigma_d\frac{1}{\omega}\max\left(\frac{\partial k}{\partial x_j}\frac{\partial \omega}{\partial x_j},0\right).
    \label{eq:omega}
\end{equation}\end{linenomath*}
where $\alpha_\omega$, $\beta_\omega$, $\sigma_\omega$, $\sigma_d$ are model coefficients.
The temperature flux dissipation term $\epsilon_{\theta i}$ is modeled based on the \textit{local isotropy} assumption as well:
\begin{linenomath*}\begin{equation}
    \epsilon_{\theta i}=0. 
\end{equation}\end{linenomath*}
The temperature variance dissipation term $\epsilon_{\theta \theta}$ is modeled using an algebraic form \begin{linenomath*}\begin{equation}
    \epsilon_{\theta \theta} = C_{\theta\theta}\frac{\varepsilon}{k}R_{\theta\theta}.
    \label{eq:epstt}
\end{equation}\end{linenomath*}
where $C_{\theta\theta}$ is another model constant.
Equations \eqref{eq:Rij}, \eqref{eq:R_theta_i}, \eqref{eq:R_theta_theta} and \eqref{eq:omega} are the eleven transport equations being solved in addition to the three RANS equations \eqref{eq:RANS_mean}. The  11-equation model contains four unclosed tensor terms, i.e., $\Phi_{ij}$, $\mathcal{D}_{p,ij}$, $\mathcal{D}_{T,ij}$, $\epsilon_{ij}$, five unclosed vector terms, i.e., $\Phi_{\theta i}$, $\mathcal{D}_{p,\theta i}$, $\mathcal{D}_{\nu,\theta i}$, $\mathcal{D}_{T,\theta i}$, $\epsilon_{\theta i}$, two unclosed scalar terms, i.e., $\mathcal{D}_{T, \theta\theta}$, $\epsilon_{\theta\theta}$, and four terms in the specific dissipation rate equation. In sum there are twenty closure coefficients. Considering this complexity, training this model such that it generalizes beyond its calibration basis is a challenging task.

\subsection{Global epistemic UQ details}
\label{sub:app-details}

The Reynolds number, Richardson number, and turbulence intensity define the FCP space.
Specifically, $Re\times Ri\times TI=[500,~10^8]\times [0,~0.1]\times [0.01,~0.2]$, covering a wide range of flow conditions.
For reference, the Reynolds number of the stratified wake behind a submarine is $Re=1 [{\rm m}]\times 10[{\rm m/s}]/10^{-6} {\rm m}^2{\rm s}^{-1}=10^7$.
We will use ${\bf F}$ to denote a point in the FCP space, and ${\bf F}_i$, $i=1$, 2, 3,..., N, to denote the sampling points in the FCP space.

The model space is 55D.
Table \ref{tab:modelspace} lists the 55 perturbations that define the model space.
The intention here is to include all possible perturbations one could incorporate in this baseline model.
Table \ref{tab:modelspace} does not include perturbations to the shape and the orientation of $\mathcal{G}_{ij}$, $\epsilon_{ij}$, $\mathcal{G}_{\theta i}$, $\epsilon_{\theta j}$, and $\epsilon_{\theta \theta}$, because they do not have a significant impact on the QoIs.
Varying a model coefficient is perturbing the magnitude of the corresponding term, and therefore we do not separately list magnitude perturbations in Table \ref{tab:modelspace}.
When perturbing a model coefficient, we vary its value by 0.5\%.
When perturbing the shape of a tensor, we vary its eigenvalues by 0.1\%.
When perturbing the orientation, we rotate the tensor about one of its eigenvectors by 0.001 rad.
Perturbation to a tensor's orientation, i.e., 0.001 rad, is small to prevent the RANS model from diverging.
Because we normalize the model's response, see Eqs. \eqref{eq:eij1} and \eqref{eq:eij2}, the amount by which we perturb the model does not affect our conclusion.
We will verify this by increasing the perturbation 10 times and compare the large and small perturbation results.
We will use ${\bf m}$ to denote a point in the model space and use $m_j$ to denote ${\bf m}$'s $j$th component.
For example, $m_1=\sigma_\omega$, and $m_{54, 55}$ describe the shape of $\mathcal{P}_{ij}$.
Because the starting point of the analysis is a lack of physical understanding, in the following, we will refrain from discussing the physics and use only model indices when referring to a specific type of model perturbation.

\begin{table}
    \caption{\label{tab:modelspace}The types of perturbations that define the model space.
    For indices 1-20, perturbations are for model coefficients.
    For indices  {21-55}, perturbations for tensor terms' orientation (Orn) and shape (Shp).}
    \begin{tabular}{|c|>{\centering\arraybackslash}m{0.03\textwidth}|
    >{\centering\arraybackslash}m{0.03\textwidth}|
    >{\centering\arraybackslash}m{0.03\textwidth}|
    >{\centering\arraybackslash}m{0.03\textwidth}|
    >{\centering\arraybackslash}m{0.03\textwidth}|
    >{\centering\arraybackslash}m{0.03\textwidth}|
    >{\centering\arraybackslash}m{0.03\textwidth}|
    >{\centering\arraybackslash}m{0.03\textwidth}|
    >{\centering\arraybackslash}m{0.03\textwidth}|
    >{\centering\arraybackslash}m{0.03\textwidth}|
    >{\centering\arraybackslash}m{0.03\textwidth}|
    >{\centering\arraybackslash}m{0.03\textwidth}|
    >{\centering\arraybackslash}m{0.03\textwidth}|}
    \hline
        Index & 1 & 2 & 3 & 4 & 5 & 6 & 7 & 8 & 9 & 10 & 11 & 12 & 13 \\
        \hline
        Perturbation & $\sigma_\omega$ & $\sigma_d$ & $\alpha_\omega$ & $\beta_\omega$ & $C_\mu$ & $C_1$ & $C_1^*$ & $C_2$ & $C_3$ & $C_3^*$ & $C_4$ & $C_5$ & $C_{bf}$ \\
        \hline
        FRSM & \multicolumn{5}{c|}{$\omega$, Eq. \eqref{eq:omega}} & \multicolumn{8}{c|}{$\Phi_{ij}$, Eq. \eqref{eq:Phiij}}\\
        \hline
    \end{tabular}
    \begin{tabular}{|c|
    >{\centering\arraybackslash}m{0.14\textwidth}|
    >{\centering\arraybackslash}m{0.14\textwidth}|
    >{\centering\arraybackslash}m{0.14\textwidth}|
    >{\centering\arraybackslash}m{0.04\textwidth}|
    >{\centering\arraybackslash}m{0.04\textwidth}|
    >{\centering\arraybackslash}m{0.04\textwidth}|
    >{\centering\arraybackslash}m{0.13\textwidth}|}
        \hline
        Index & 14 & 15 & 16 & 17 & 18 & 19 & 20 \\
        \hline
        Perturbation & $D$ & $Pr_{tur,\theta i}$ & $Pr_{tur,\theta \theta}$ & $C_{\theta 1}$ & $C_{\theta 2}$ & $C_{\theta 3}$ & $C_{\theta \theta}$ \\
        \hline
        FRSM & $\mathcal{D}_{ij}$, Eq. \eqref{eq:Dij} & $\mathcal{D}_{\theta i}$, Eq. \eqref{eq:Dti} & $\mathcal{D}_{\theta \theta}$, Eq. \eqref{eq:Dtt} & \multicolumn{3}{c|}{ $\Phi_{\theta i}$, Eq. \eqref{eq:Phiti}} & $\epsilon_{\theta\theta}$,  Eq. \eqref{eq:epstt} \\
        \hline
    \end{tabular}
    \begin{tabular}{|c|>{\centering\arraybackslash}m{0.06\textwidth}|
    >{\centering\arraybackslash}m{0.07\textwidth}|
    >{\centering\arraybackslash}m{0.06\textwidth}|
    >{\centering\arraybackslash}m{0.07\textwidth}|
    >{\centering\arraybackslash}m{0.06\textwidth}|
    >{\centering\arraybackslash}m{0.07\textwidth}|
    >{\centering\arraybackslash}m{0.06\textwidth}|
    >{\centering\arraybackslash}m{0.07\textwidth}|}
        \hline
        Index & 21-23 & 24, 25 & 26-28 & 29, 30 & 31-33 & 34, 35 & 36-38 & 39,40 \\
        \hline
        Perturbation & Orn & Shp & Orn & Shp & Orn & Shp & Orn & Shp \\
        \hline
        {FRSM, Eq. \eqref{eq:Phiij}} & \multicolumn{2}{c|}{1st term} &  \multicolumn{2}{c|}{2nd term} & \multicolumn{2}{c|}{3rd term} & \multicolumn{2}{c|}{4th term} \\
        \hline
    \end{tabular}
    \begin{tabular}{|c|>{\centering\arraybackslash}m{0.06\textwidth}|
    >{\centering\arraybackslash}m{0.07\textwidth}|
    >{\centering\arraybackslash}m{0.09\textwidth}|
    >{\centering\arraybackslash}m{0.09\textwidth}|
    >{\centering\arraybackslash}m{0.07\textwidth}|
    >{\centering\arraybackslash}m{0.07\textwidth}|}
        \hline
        Index & 41-43 & 44, 45 & 46-48 & 49, 50 & 51-53 & 54, 55 \\
        \hline
        Perturbation & Orn & Shp & Orn & Shp & Orn & Shp \\
        \hline
        FRSM & \multicolumn{2}{c|}{5th term} & \multicolumn{2}{c|}{6th term in Eq. \eqref{eq:Phiij}} & \multicolumn{2}{c|}{ $\mathcal{P}_{ij}$ in Eq. \eqref{eq:Rij}} \\
        \hline
    \end{tabular}
    
\end{table}

We have 9 QoIs, namely, the momentum thickness $\delta_\theta$, the mean velocity $U$, the mean temperature $T$ (equivalent to the mean density), the mean kinetic energy MKE, the second order moments $\left<u'u'\right>$, $\left<v'v'\right>$, $\left<u'v'\right>$, $\left<v'T'\right>$, and $\left<T'T'\right>$.
These QoIs are functions of ${\bf M}$ and ${\bf F}$. 
Except for the momentum thickness $\delta_\theta$, which is a function of $t$ not $y$, all other 8 QoIs are functions of both $y$ and $t$.
Evaluating $e_{i,j,{\rm QoI}}$ for a given QoI requires two evaluations of the FRSM, one at ${\bf M}_o$ and one at ${\bf M}_o+\delta_j$:
\begin{linenomath*}\begin{equation}
    e_{i,j,\delta_\theta}=\sqrt{\frac{1}{t_e} \int_{0}^{t_e}dt\left[\frac{{\delta_\theta}({\bf F}_i,{\bf M}_o+\delta_j)-{\rm \delta_\theta}({\bf F}_i,{\bf M}_o)}{\delta_j}\right]^2},
    \label{eq:eij1}
\end{equation}\end{linenomath*}
for the QoI $\delta_\theta$, and
\begin{linenomath*}\begin{equation}
    e_{i,j,{\rm QoI}}=\sqrt{\frac{1}{t_e\delta_{\omega,0}}\int_{-\infty}^{\infty}dy \int_{0}^{t_e}dt\left[\frac{{\rm QoI}({\bf F}_i,{\bf M}_o+\delta_j)-{\rm QoI}({\bf F}_i,{\bf M}_o)}{\delta_j}\right]^2},
    \label{eq:eij2}
\end{equation}\end{linenomath*}
for the other 8 QoIs.
Here, ${\bf M}_o$ is the original model and $\delta_j$ is the perturbation in the $j$th dimension in the model space
It then follows from Eq. \eqref{eq:sigma-mu} that $\sigma_{j,{\rm QoI}}$ and $\mu_{j,{\rm QoI}}$ are
\begin{linenomath*}\begin{equation}
    \mu_{j,{\rm QoI}}=\frac{1}{N}\sum_{i=1}^N e_{i,j,{\rm QoI}},~~\sigma_{j,{\rm QoI}}^2=\frac{1}{N}\sum_{i=1}^N (e_{i,j,{\rm QoI}}-\mu_{j,{\rm QoI}})^2.
    \label{eq:sigma-mu2}
\end{equation}\end{linenomath*}
For a given QoI and a give type of perturbation $j$, $\mu_{j,{\rm QoI}}$ and $\sigma_{j,{\rm QoI}}$ are two numbers.

\subsection{Global epistemic UQ results}
\label{sub:app-results}

\subsubsection{The number of samplings in the flow controlling parameter space}
\label{subsub:N}

\begin{figure}
    \centering
    \includegraphics[width=0.45\textwidth]{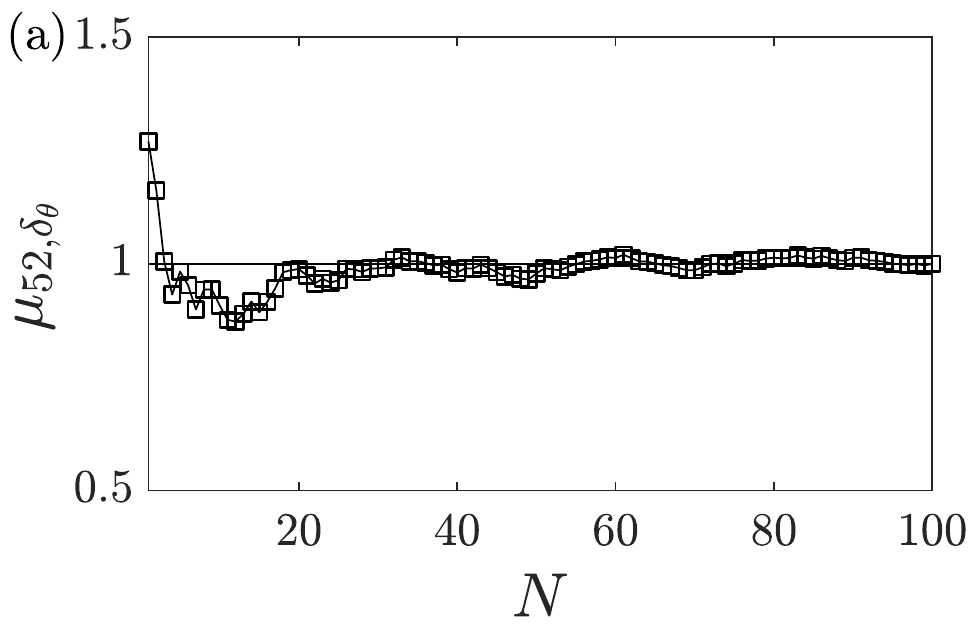}~~~
    \includegraphics[width=0.45\textwidth]{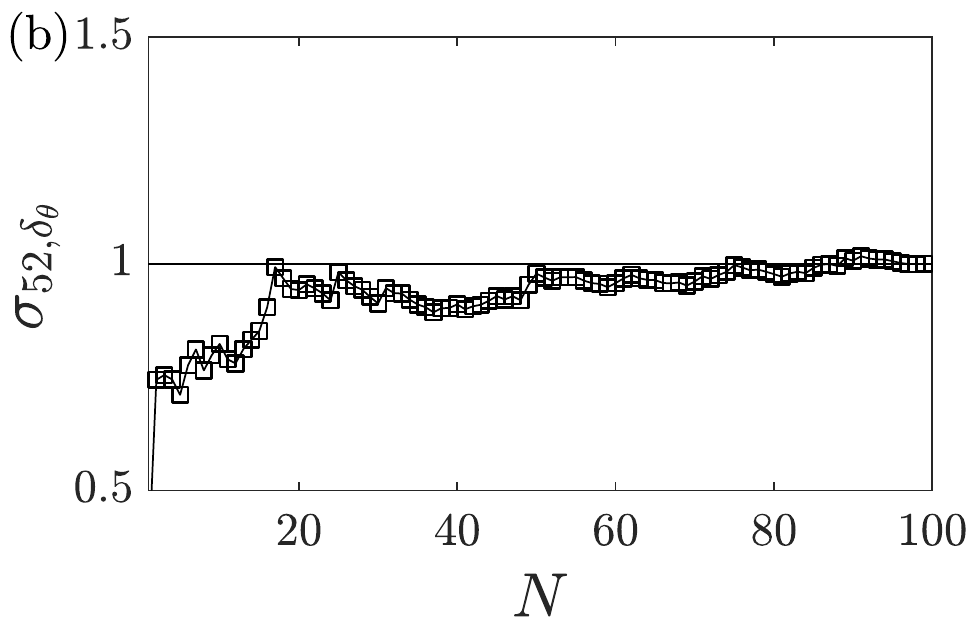}
    \caption{(a) $\mu_{52,\delta_\theta}$ and (b) $\sigma_{52,\delta_\theta}$ as a function of the sampling points $N$ in the full FCP space. 
    The results are normalized by their respective values at $N=100$.
    }
    \label{fig:N}
\end{figure}

To obtain reliable estimates of $\sigma_{j,{\rm QoI}}$ and $\mu_{j,{\rm QoI}}$, we must sufficiently sample the FCP space.
Figure \ref{fig:N} shows $\sigma_{52,\delta_\theta}$ and $\mu_{52,\delta_\theta}$ as a function of the number of sampling $N$.
$e_{i,52,\delta_\theta}$ varies significantly within the FCP space, and $\sigma_{52,\delta_\theta}$, $\mu_{52,\delta_\theta}$ is in the H.E., H.I quadrant (see section \ref{subsub:3D}).
Estimating $\sigma_{52,\delta_\theta}$, $\mu_{52,\delta_\theta}$ is therefore more difficult than, say, estimating $\sigma$, $\mu$ in the H.E., L.I. quadrant.
From figure \ref{fig:N}, we observe that $N=100$ returns reliable estimates of $\sigma$, $\mu$. 
By sampling 100 times in the FCP space, each global epistemic UQ requires $N\times N'=100\times 56=5,600$ evaluations of the RANS model.

\subsubsection{Global epistemic UQ in a sub flow controlling parameter space}
\label{subsub:2D}

We first consider shear layers with no thermal stratification.
The FCP space reduces to 2D, i.e., $Re\times TI=[500,~10^8]\times [0.01,~0.2]$ with $Ri=0$, and the temperature reduces to a passive scalar.
The flow physics of an unstratified shear layer is considerably simpler than a stratified one, and therefore we expect low inconsistency for perturbations in the model space.

\begin{figure}
    \centering
    \includegraphics[height=0.162\textwidth]{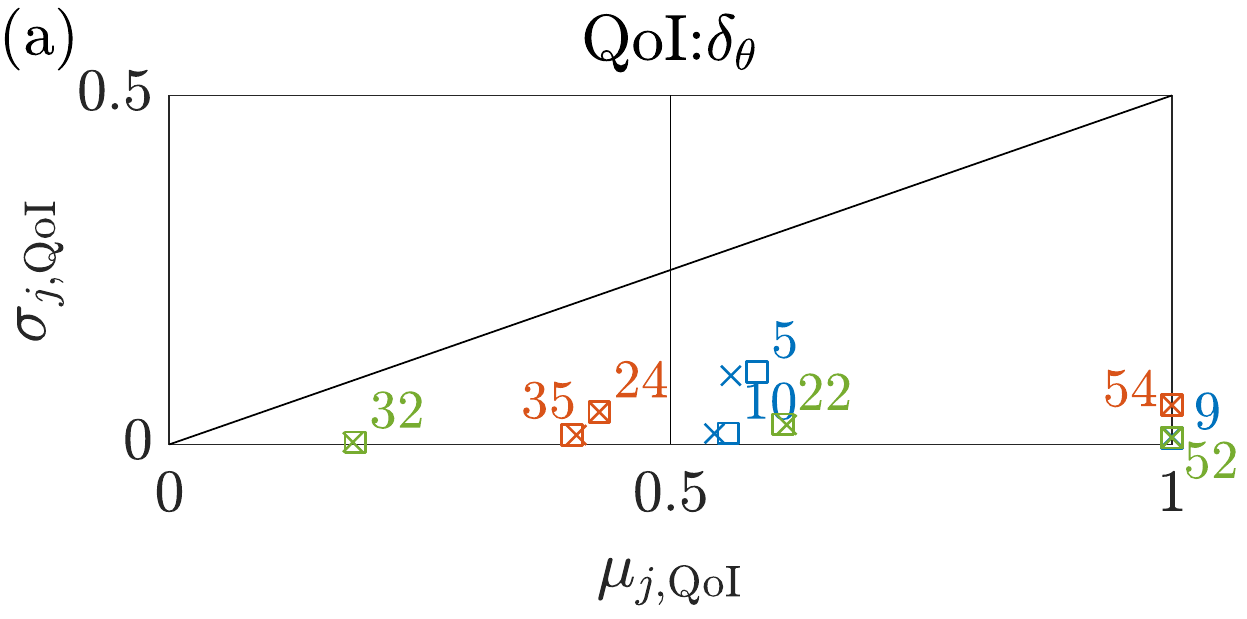}
    \includegraphics[height=0.162\textwidth]{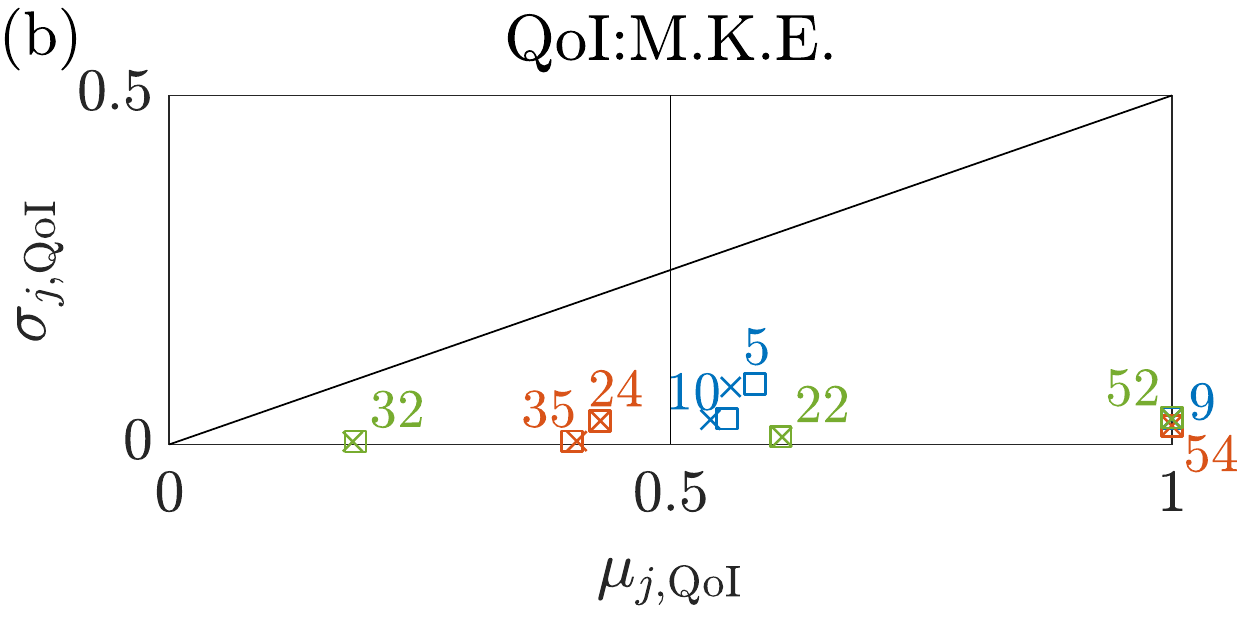}
    \includegraphics[height=0.162\textwidth]{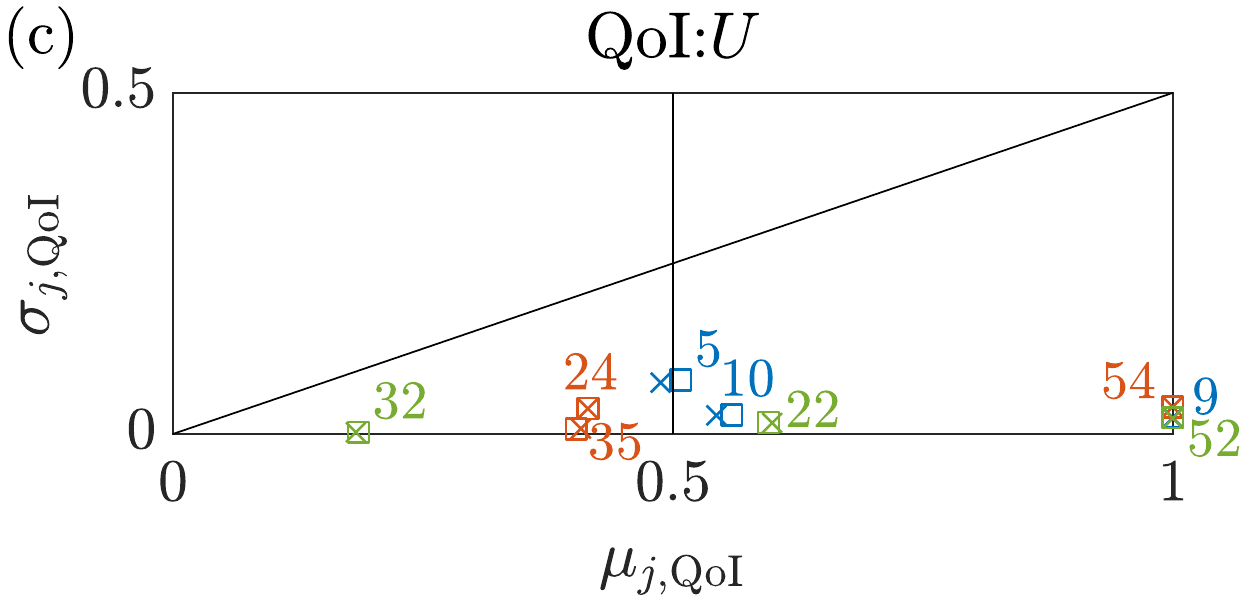}\\
    \includegraphics[height=0.160\textwidth]{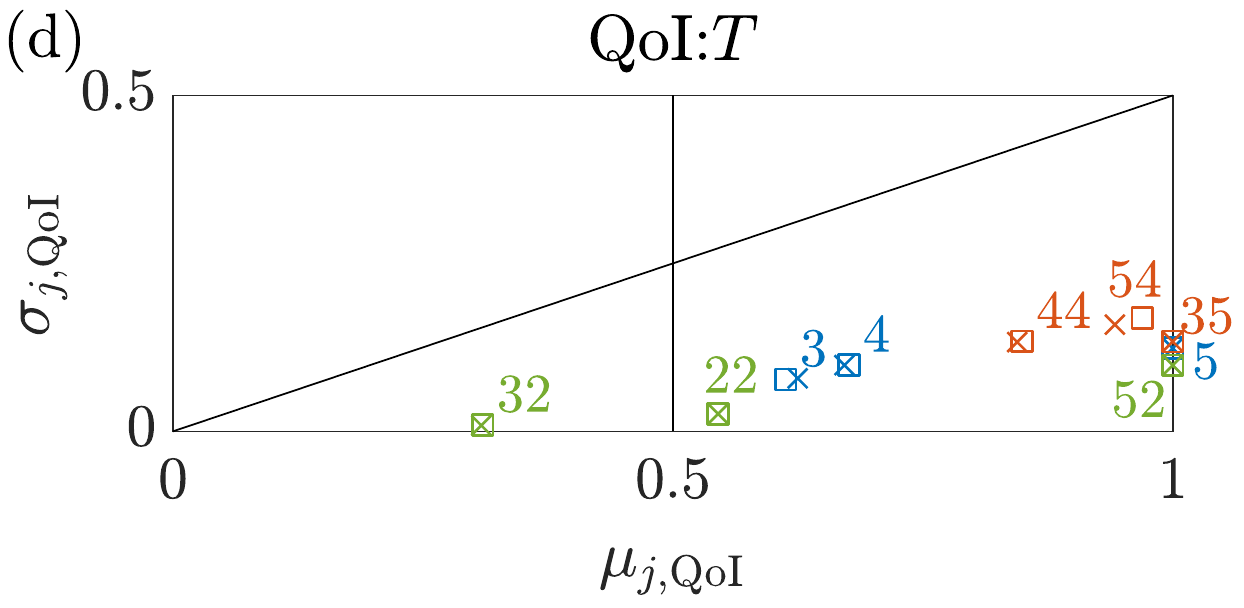}
    \includegraphics[height=0.160\textwidth]{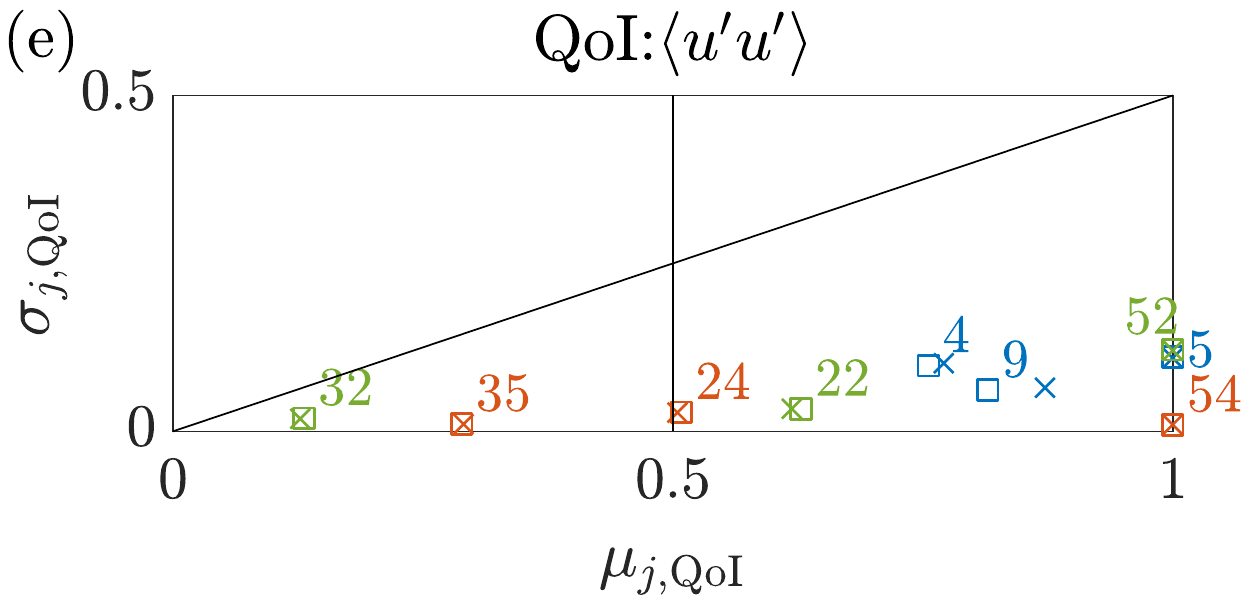}
    \includegraphics[height=0.160\textwidth]{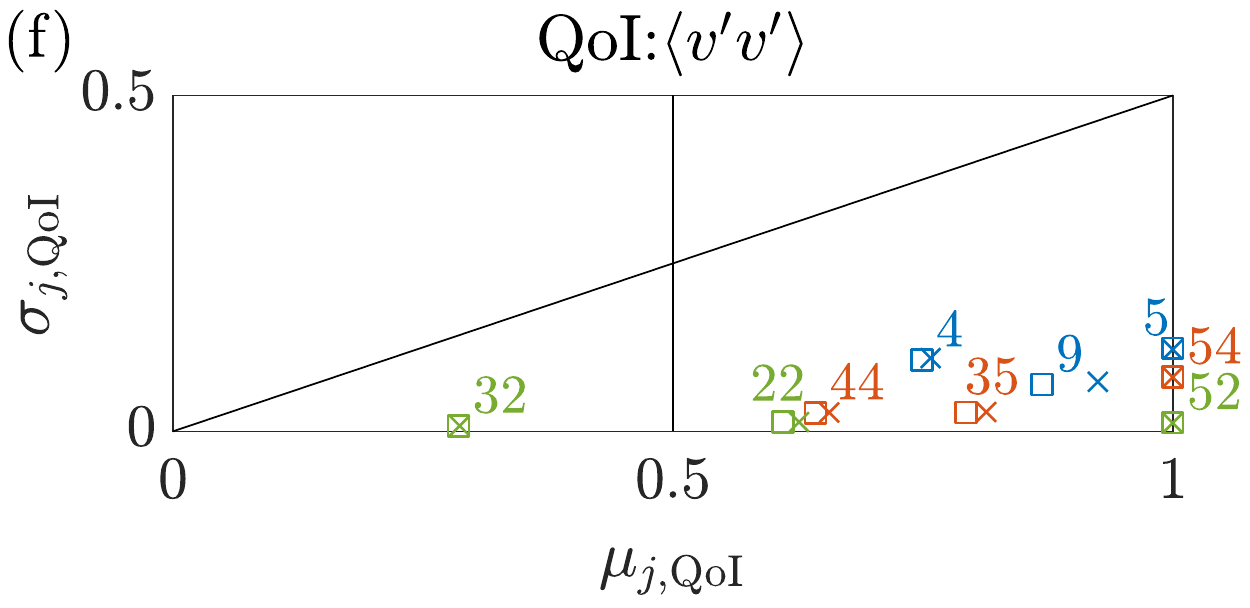}\\
    \includegraphics[height=0.160\textwidth]{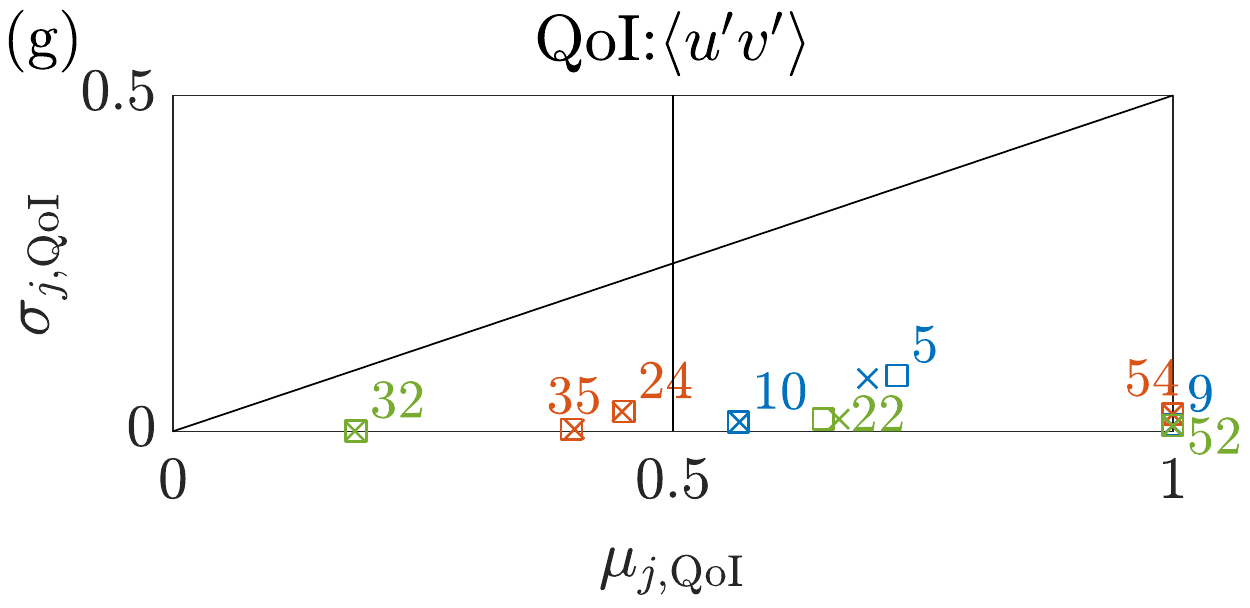}
    \includegraphics[height=0.160\textwidth]{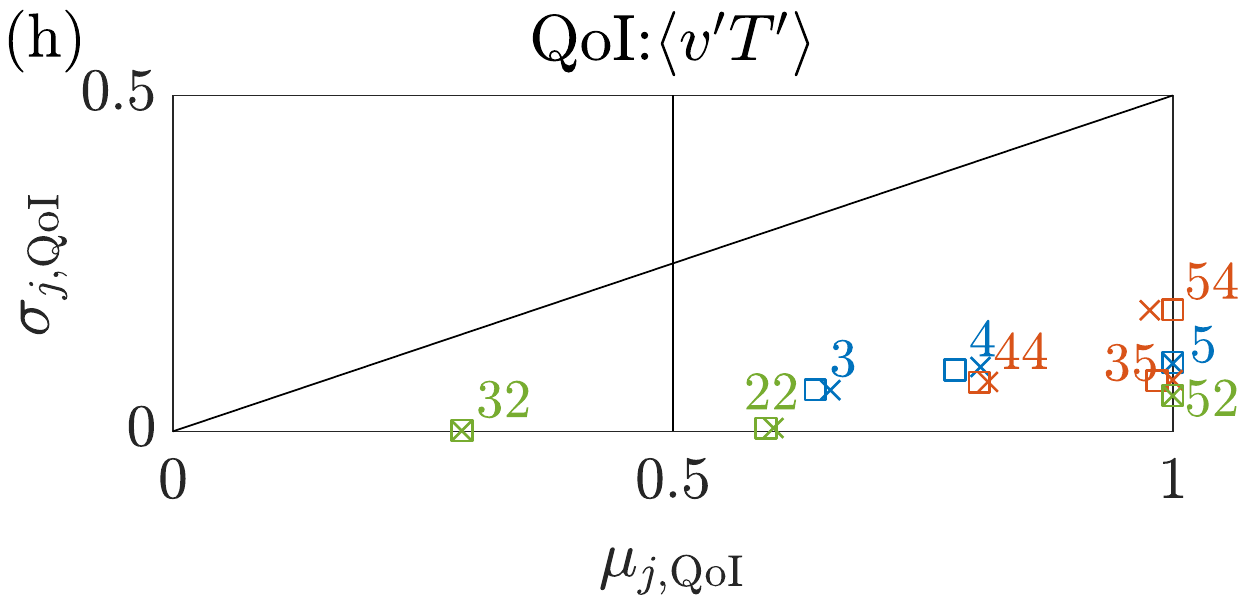}
    \includegraphics[height=0.160\textwidth]{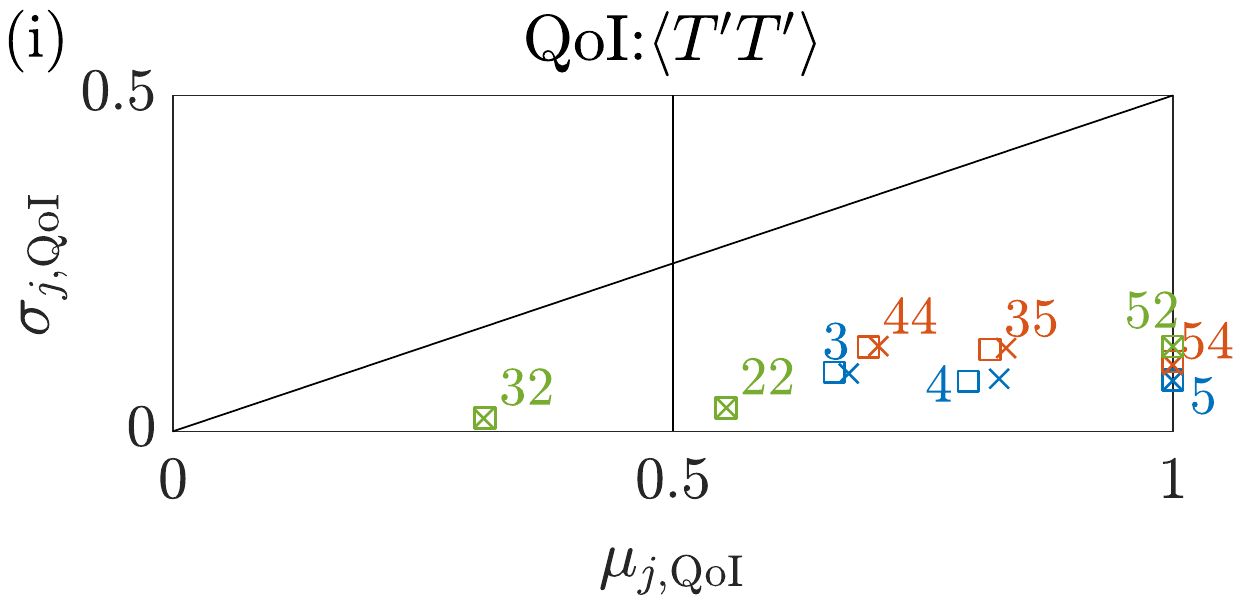}\\
    \caption{$\sigma-\mu$ diagrams for unstratified shear layer. 
    For visualization purposes, we show $\sigma$, $\mu$ for the 3 most effective shape perturbations, the 3 most effective orientation perturbations, and the 3 most effective model-coefficient perturbations.
    Perturbation types are color-coded.
    We use green for orientation perturbations, red for shape perturbations, and blue for model-coefficient perturbations.
    For a given QoI and a given type of model perturbation, the $\sigma$, $\mu$ values are normalized with $\max_j[\sigma_j]$ among the given type of model perturbation (shape/orientation/model-coefficient).
    Square symbols are results due to small perturbations, and cross symbols are due to large perturbations.
    }
    \label{fig:Results-S1}
\end{figure}

We sample the 2D FCP space 100 times.
Figure \ref{fig:Results-S1} shows $\mu_{j,{\rm QoI}}$ and $\sigma_{j,{\rm QoI}}$ for all QoIs.
For visualization purposes, we show the $\sigma$ and $\mu$ values for the 3 most effective shape perturbations, the 3 most effective orientation perturbations, and the 3 most effective model coefficient perturbations.
For each QoI and a type of perturbation (shape perturbation, orientation perturbation, or model coefficient perturbation), we normalize $\mu_{j,{\rm QoI}}$ and $\sigma_{j,{\rm QoI}}$ with $\max_j[\mu_{j,{\rm QoI}}]$.
We make the following observations.
First, large perturbations and small perturbations lead to similar results.
Second, we see that all perturbations in figure \ref{fig:Results-S1} have low inconsistency.
The effective perturbations are fairly consistent between different QoIs, i.e., a perturbation that is effective for one QoI is also effective for other QoIs.
Perturbations 22, 32, and 52 are the most effective orientation perturbation for all 9 QoIs.
Perturbations 35 and 54 are among the 3 most effective shape perturbation for all 9 QoIs.
Perturbation 9 is the most effective model coefficient perturbation for all velocity QoIs. 
Perturbation 5 is the most effective model cofficint perturbation for temperature QoIs, i.e., $\left<T\right>$, $\left<v'T'\right>$, and $\left<T'T'\right>$.
Despite the lack of physical arguments guiding this analysis, the fact that perturbations 52 and 54, i.e., perturbing the production term in the Reynolds transport equation, are effective, is consistent with the previous work \cite{emory2013modeling,gorle2019epistemic}.
This also serves as a validation of the proposed global epistemic UQ analysis.

From a model calibration standpoint, global epistemic uncertainty UQ help us to identify the most effective perturbations for model calibration.
Because the most effective perturbations are also consistent, we conclude from figure \ref{fig:Results-S1}, {\it a priori}, that for any one of the 9 QoIs, calibrating perturbations 52, 54, 9, and 5 will lead to improvements at other flow conditions in the FCP.
Also, because the effective perturbations are consistent between the 9 QoIs, we can conclude, {\it a priori}, that if one perturbs 52, 54, 9, and 5, an improvement for one QoI would lead to an improvement in other QoIs.

For a FRSM, the production term $\mathcal{P}_{ij}$ is closed, and therefore calibrating the production term (i.e., perturbations 52 and 54) is not a good idea.
Considering that the physics is such that the production and the dissipation approximately balance \cite{brucker2007evolution}, the results suggest that we need a more accurate model for the dissipation term.
This makes sense: one uses FRSM to handle Reynolds stress's anisotropy, but the present model for the dissipation term $\varepsilon_{ij}$ is itself isotropic, which obviously requires improvements.
Also, calibrating perturbation 5 may not be practical -- albeit being an effective and consistent perturbation.
Perturbation 5 is a model coefficient perturbation. 
It corresponds to the Prandtl-Kolmogorov constant $C_\mu$.
While there is no guarantee that the Prandtl-Kolmogorov constant is universal, its value $C_\mu=0.09$ works well for many flows, and varying it will negatively impact the model's robustness. (Note that the global epistemic UQ does not ``know'' this because the other flows are not part of this analysis.)
In all, model calibration is a multi-faceted problem.
While effectiveness and consistency are two important aspects of the problem, model calibration must also account for, e.g., robustness and prior knowledge, which are not accounted for in our global epistemic UQ analysis.
As a result, we cannot exactly follow results of our global epistemic UQ analysis when calibrating a model.

\subsubsection{Global epistemic UQ in the full flow controlling parameter space}
\label{subsub:3D}

\begin{figure}
    \centering
    \includegraphics[height=0.162\textwidth]{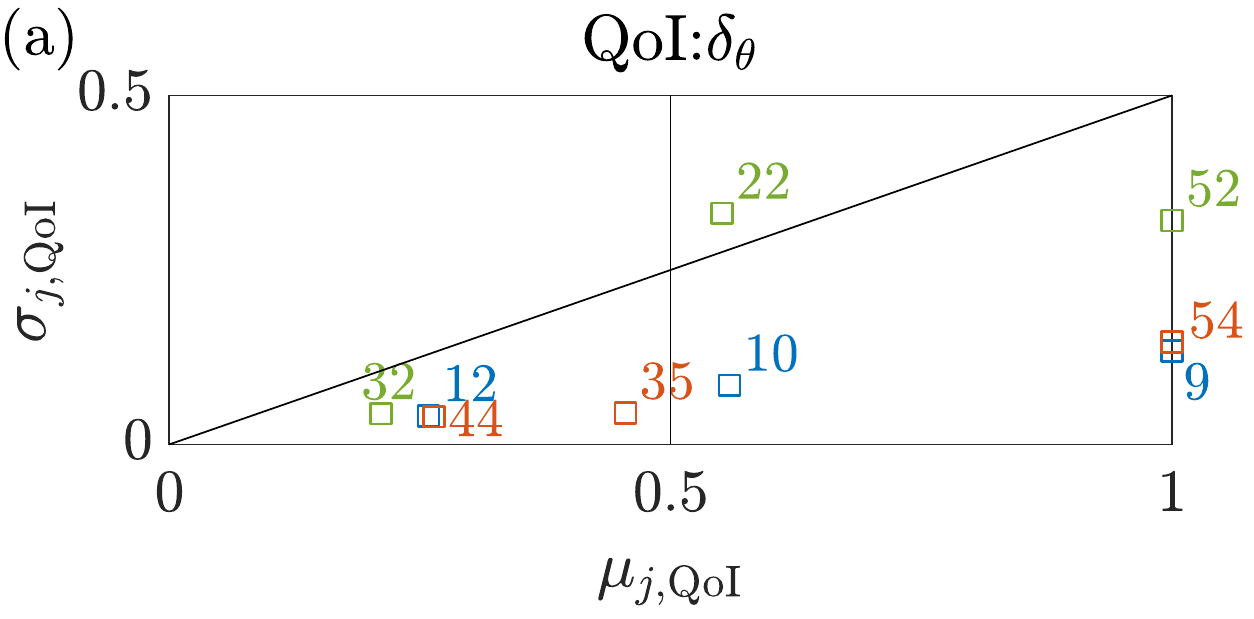}
    \includegraphics[height=0.162\textwidth]{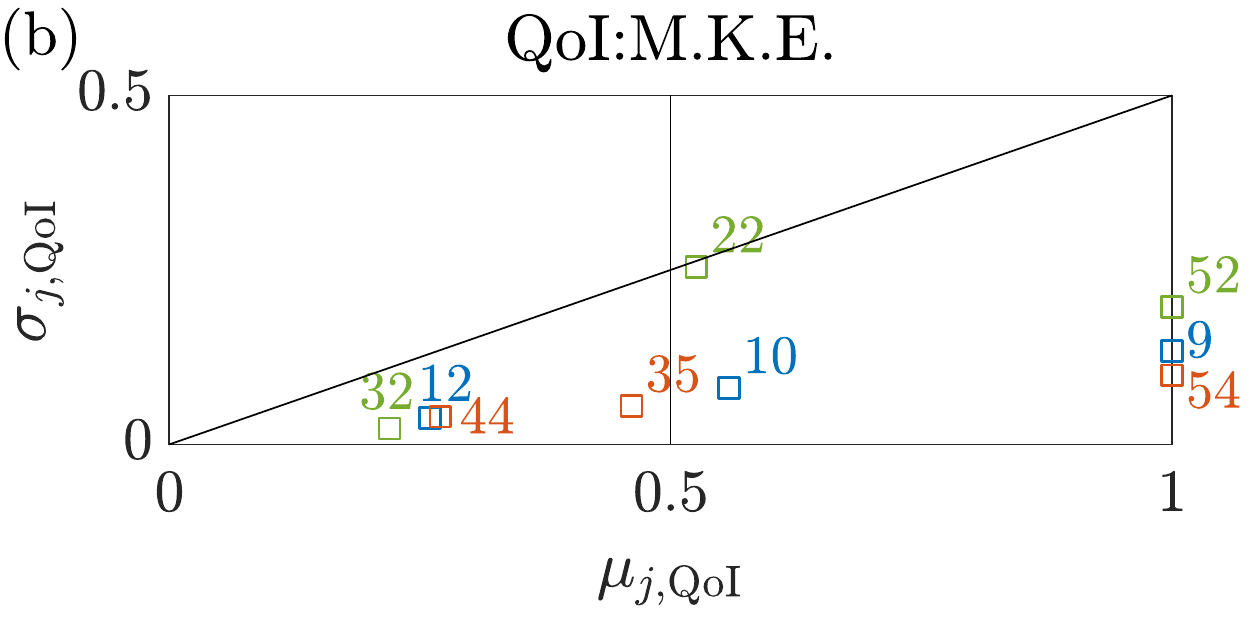}
    \includegraphics[height=0.162\textwidth]{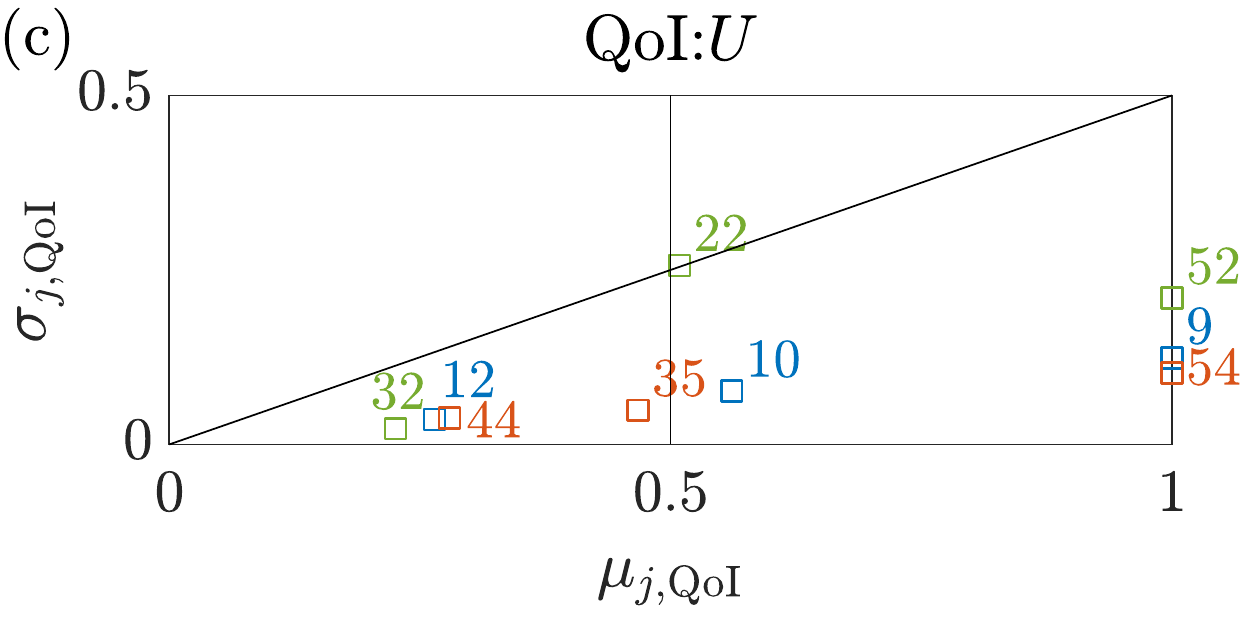}\\
    \includegraphics[height=0.161\textwidth]{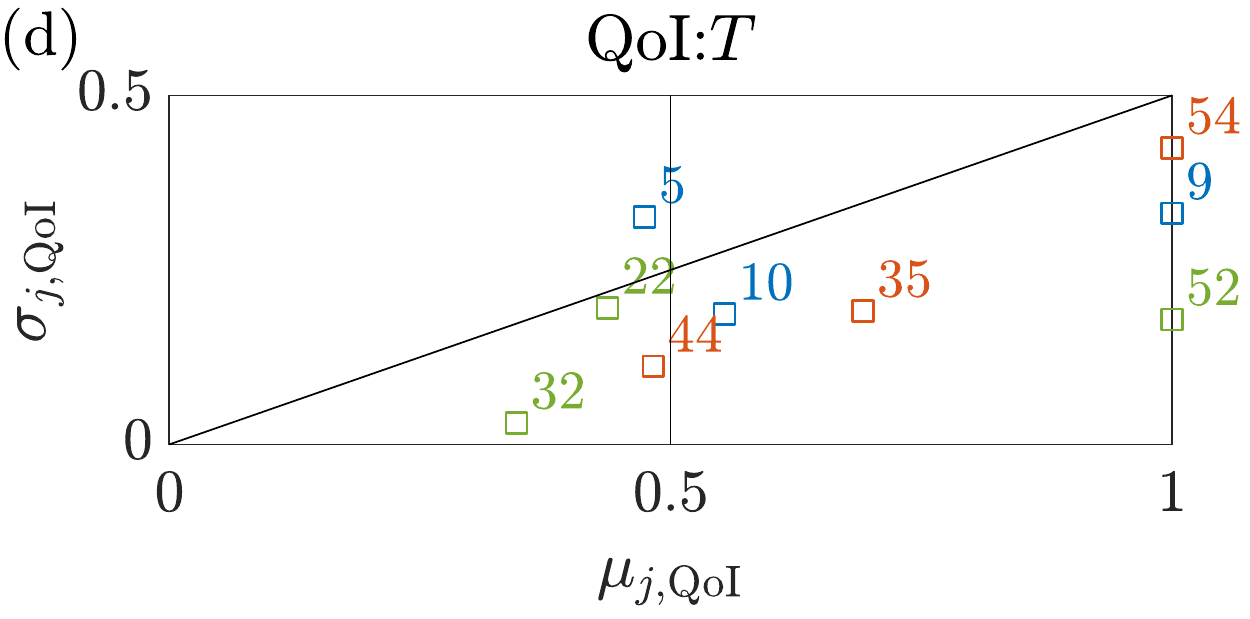}
    \includegraphics[height=0.161\textwidth]{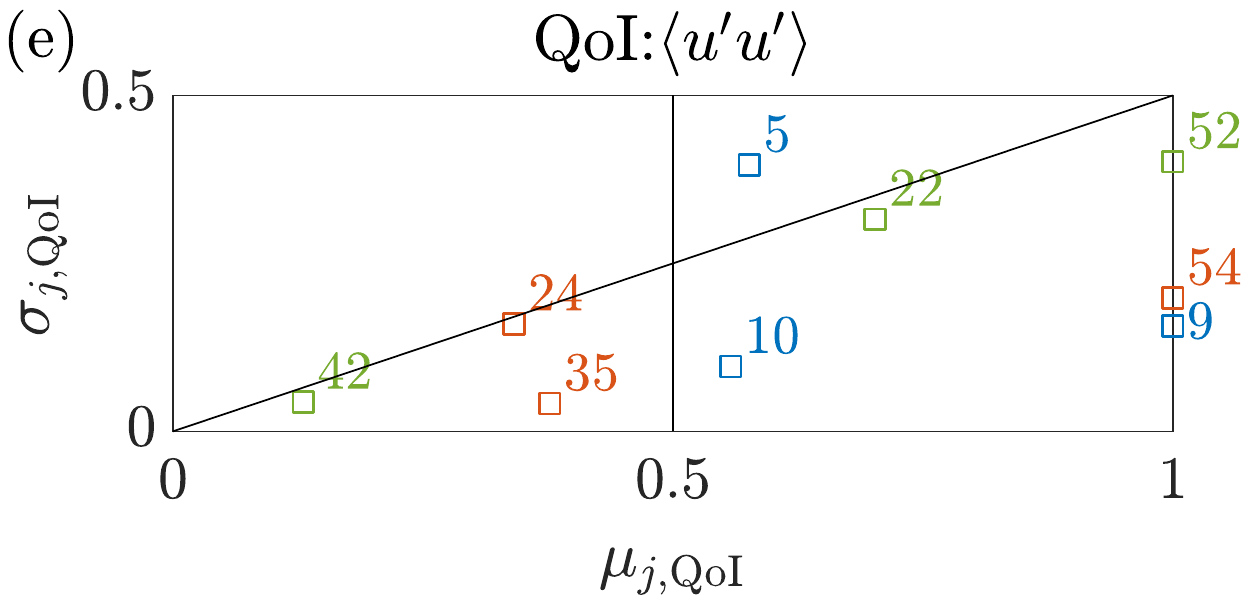}
    \includegraphics[height=0.161\textwidth]{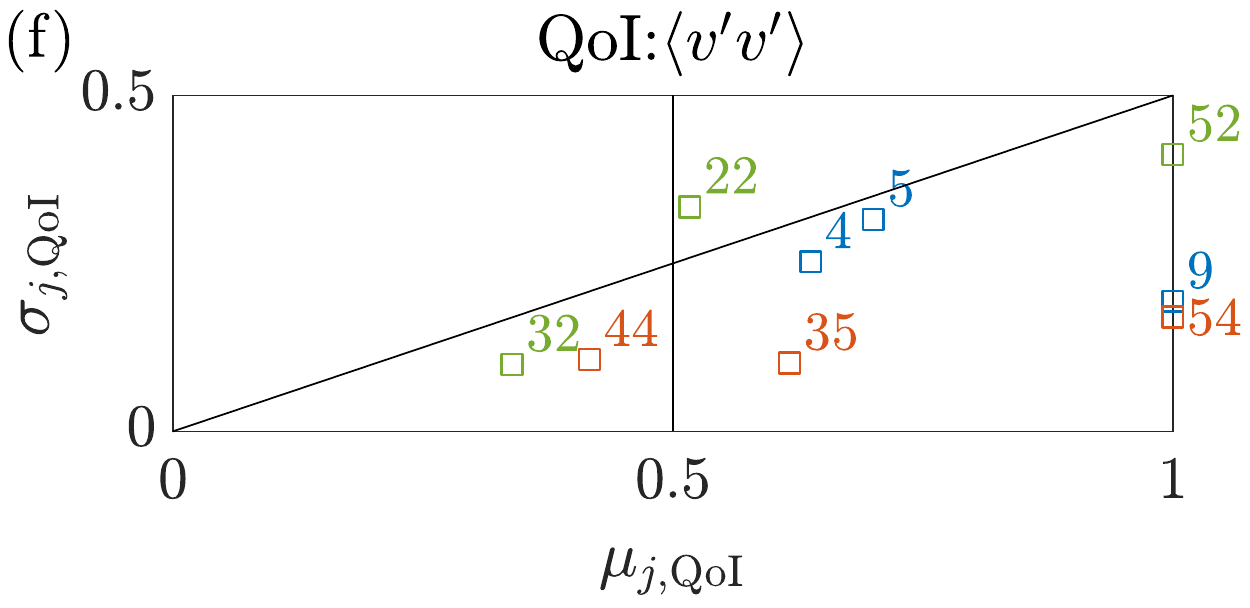}\\
    \includegraphics[height=0.162\textwidth]{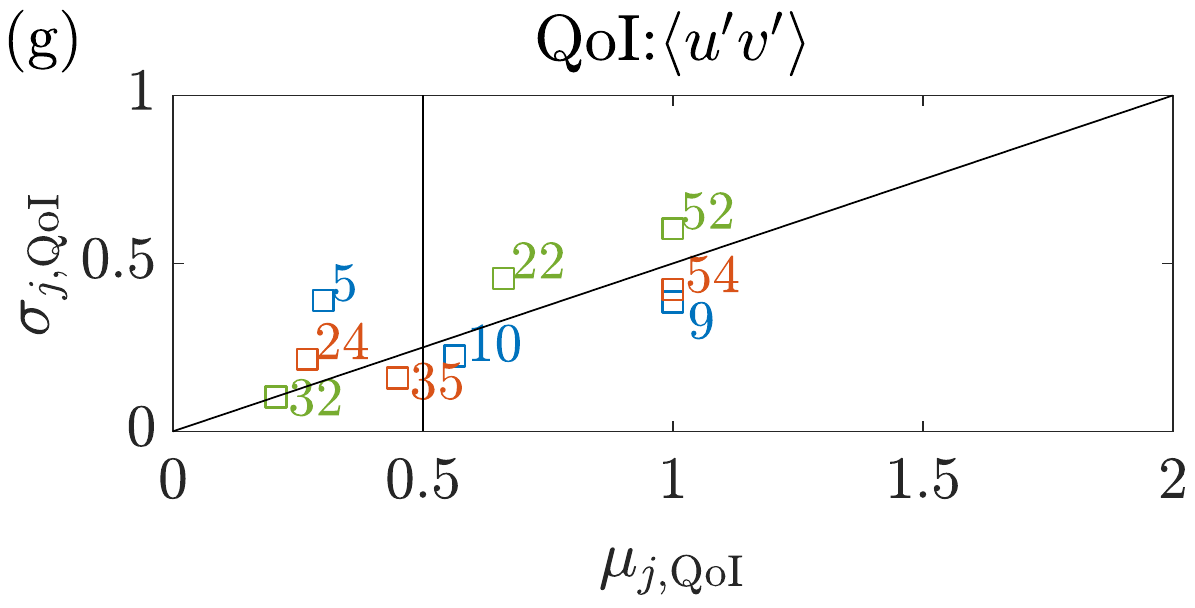}
    \includegraphics[height=0.162\textwidth]{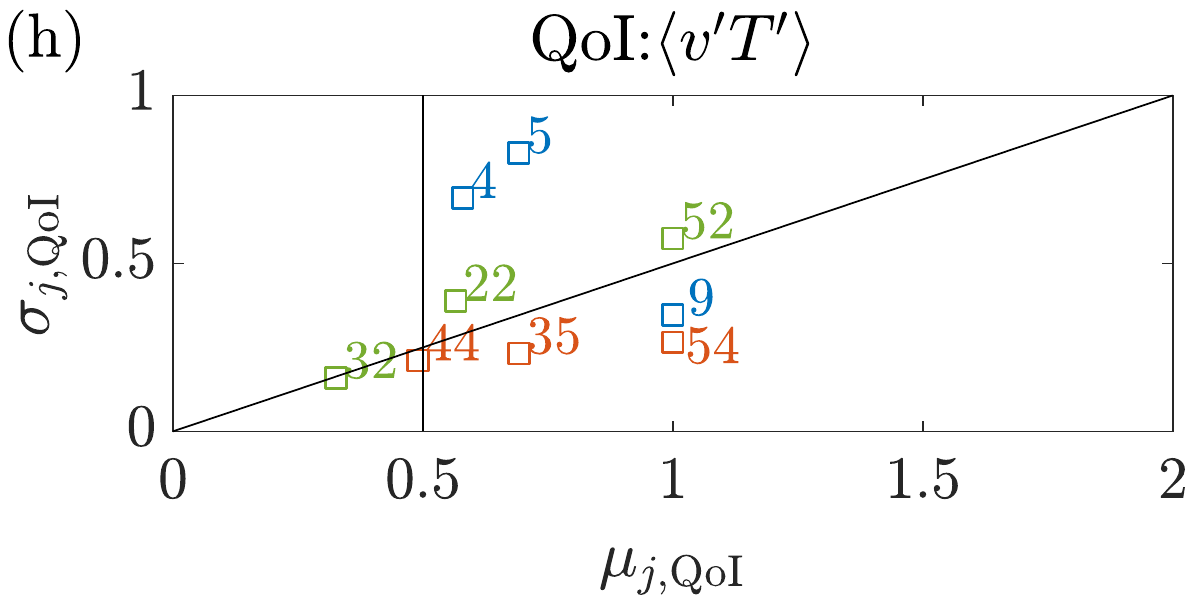}
    \includegraphics[height=0.162\textwidth]{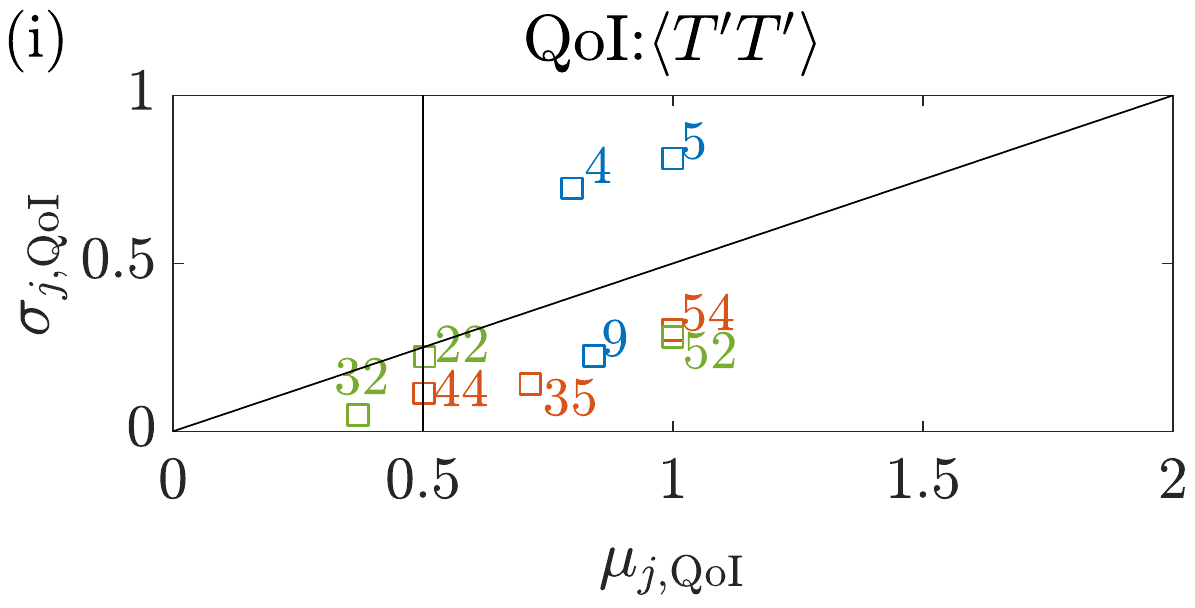}\\
    
    \caption{$\sigma-\mu$ diagram for stratified shear layer. 
    The legends are the same as in figure \ref{fig:Results-S1}.}
    \label{fig:Results-S3}
\end{figure}

The flow physics of a stratified shear layer is considerably more complicated than an unstratified one.
Because different physical processes dominate in different flow regimes, we expect that model perturbations have high inconsistency.
Figure \ref{fig:Results-S3} shows the effectiveness and inconsistency of the most effective perturbations.
For visualization purposes, again, we show the 3 most effective orientation perturbations, the 3 most effective shape perturbations, and the 3 most effective model coefficient perturbations.
We see that many of the effective perturbations have fairly high inconsistency: they are in the H.I. quadrants or very close to the H.I. quadrants.
Nonetheless, the effective perturbations are still consistent between different QoIs. 
Perturbations 9, 52, and 54 are among the most effective perturbations for all QoIs.
Perturbations 35, 44, 54 are the 3 most effective shape perturbations for all but $\left<u'u'\right>$ and $\left<u'v'\right>$, for which, perturbation 24 replaces perturbation 44 as one of the 3 most effective shape perturbations.
Perturbations 22, 32, and 52 are the 3 most effective orientation perturbations for all but $\left<u'u'\right>$ for which 42 replaces 22 as one of the 3 most effective shape perturbations.
Perturbation 9 is among the 3 most effective model perturbations for all 9 QoIs. 
Perturbation 10 is among the 3 most effective model coefficient perturbations for all but $\left<v'v'\right>$, $\left<v'T'\right>$, and $\left<T'T'\right>$. 
Comparing the results in this subsection and section \ref{subsub:2D}, the most effective perturbations are the same for unstratified and stratified flows.
Perturbations of the production term are still among the most effective model perturbations.

\begin{figure}
    \centering
    \includegraphics[width=0.32\textwidth]{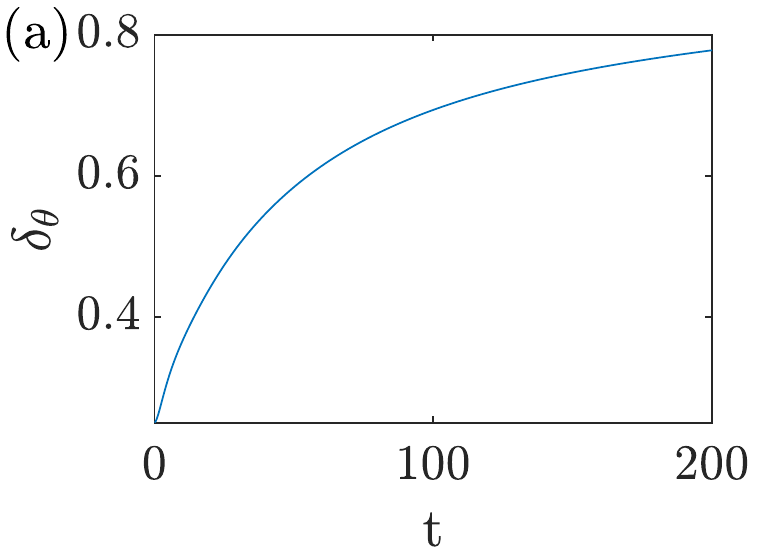}
    \includegraphics[width=0.32\textwidth]{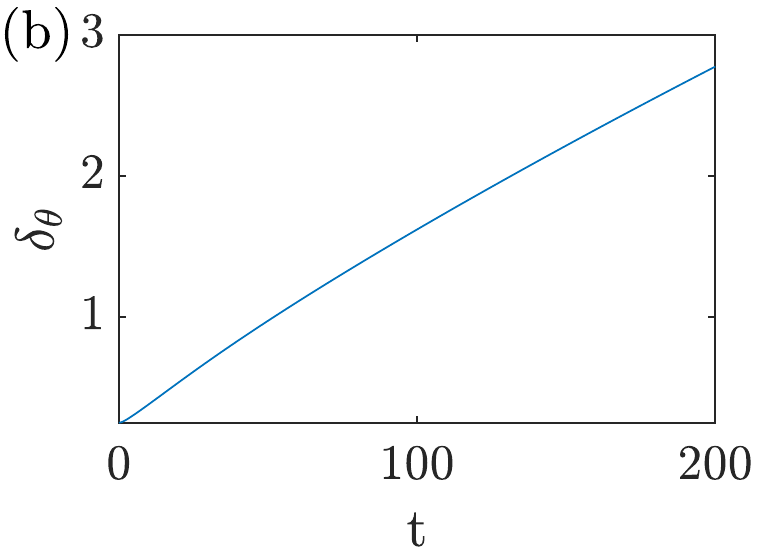}
    \includegraphics[width=0.32\textwidth]{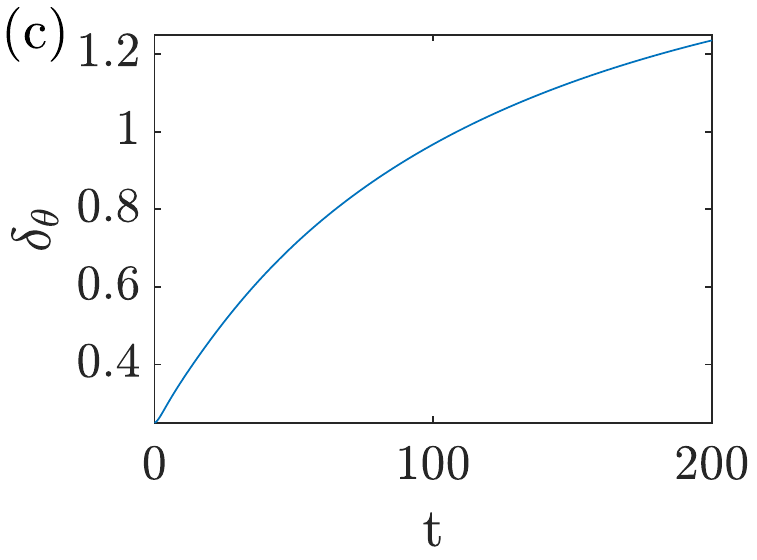}
    \caption{The FRSM predicted time evolution of the shear layer thickness $\delta_\theta$ at three flow conditions in the FCP space.
    The exact $Re$, $TI$, and $Ri$ are not important.
    The shear layer thickness is normalized with its value at $t=0$.
    (a) The shear layer has entered the late stage at $t=200$.
    (b) The shear layer remains at the early stage at $t=200$. 
    (c) The shear layer is somewhat ambiguous in terms of whether it is at the early stage or late stage at $t=200$. }
    \label{fig:Results-S2-delta}
\end{figure}

From a model calibration standpoint, high inconsistency is undesirable.
If a type of model perturbation has high inconsistency, calibrating the corresponding term/coefficient is less likely to lead to improvements that generalize.
In that case, the physics is such that one needs more than one model, or more than one variant of the baseline model, for the flows in the FCP space.
For a stratified shear layer, the different physics is the physics of the flow in the early stage and the late stage, where the buoyancy and the turbulent convection play the dominant role, respectively.
Figure \ref{fig:Results-S2-delta} shows the FRSM predicted $\delta_\theta(t)$ as a function of time at three flow conditions in the FCP space. 
The flow in figure \ref{fig:Results-S2-delta} (a) is still at the early stage at $t=200$, and the flow in figure \ref{fig:Results-S2-delta} (b) is at the late stage at $t=200$.
Expecting an improvement that works for an early-stage shear layer to also work for a late-stage shear layer would be hard, and this had resulted in the high inconsistency in figure \ref{fig:Results-S3}.

\begin{figure}
    \centering
    \includegraphics[height=0.162\textwidth]{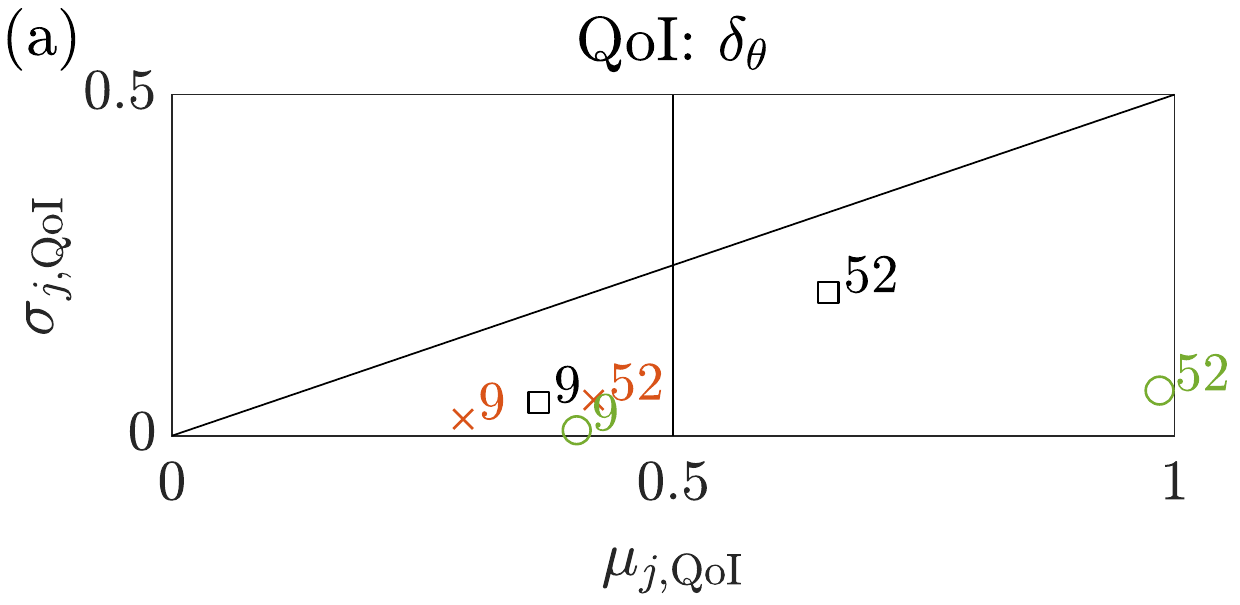}
    \includegraphics[height=0.162\textwidth]{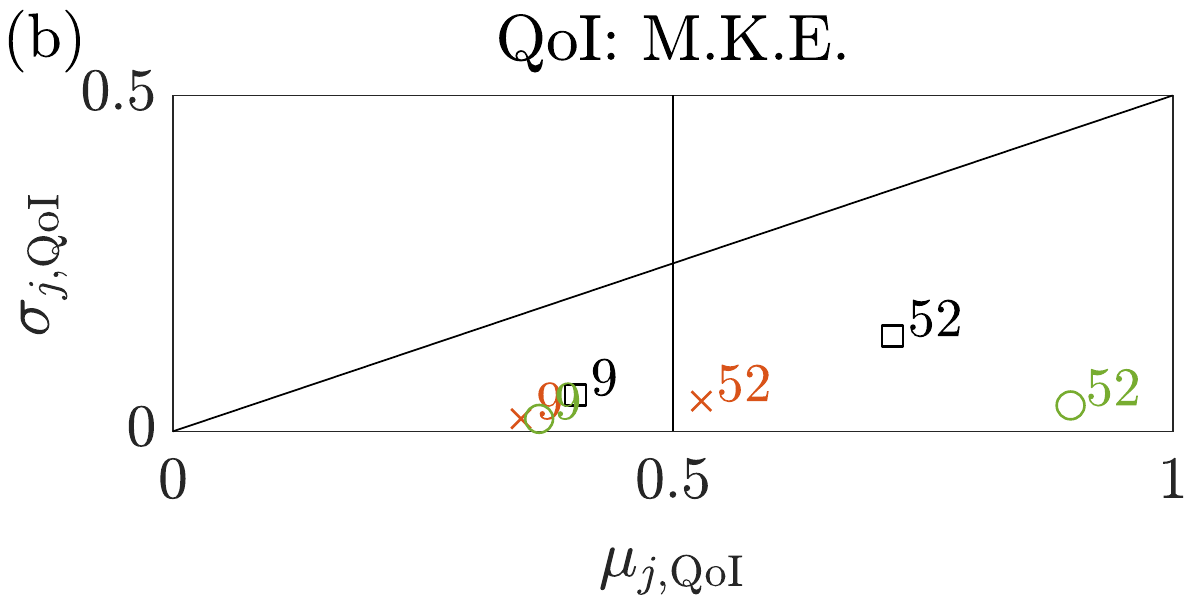}
    \includegraphics[height=0.162\textwidth]{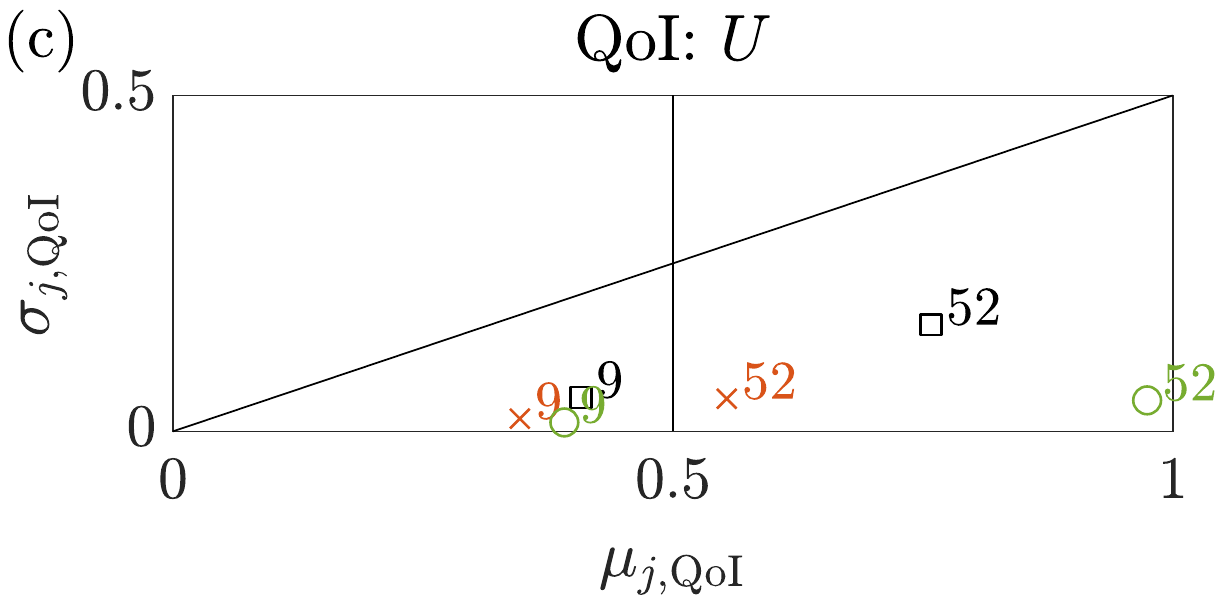}\\
    \includegraphics[height=0.162\textwidth]{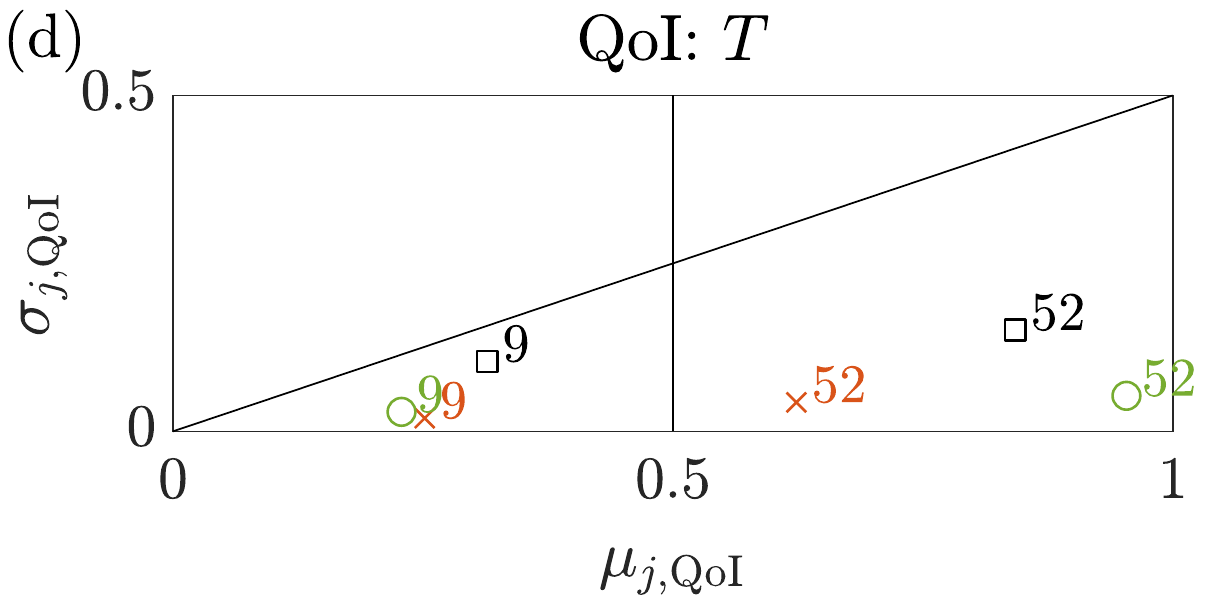}
    \includegraphics[height=0.162\textwidth]{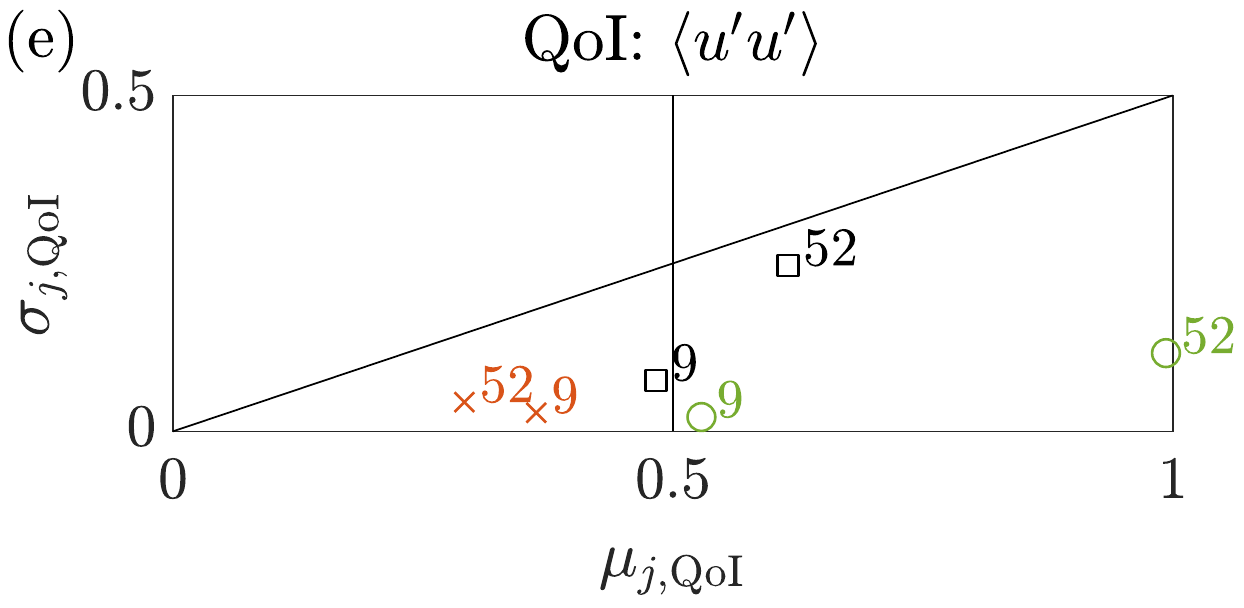}
    \includegraphics[height=0.162\textwidth]{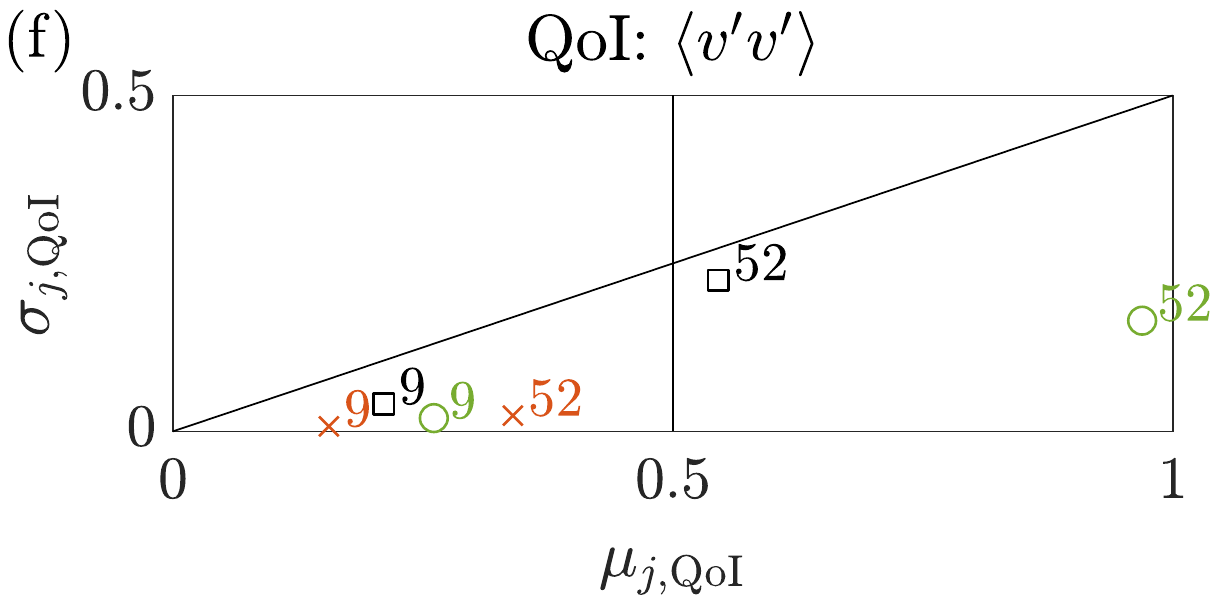}\\
    \includegraphics[height=0.162\textwidth]{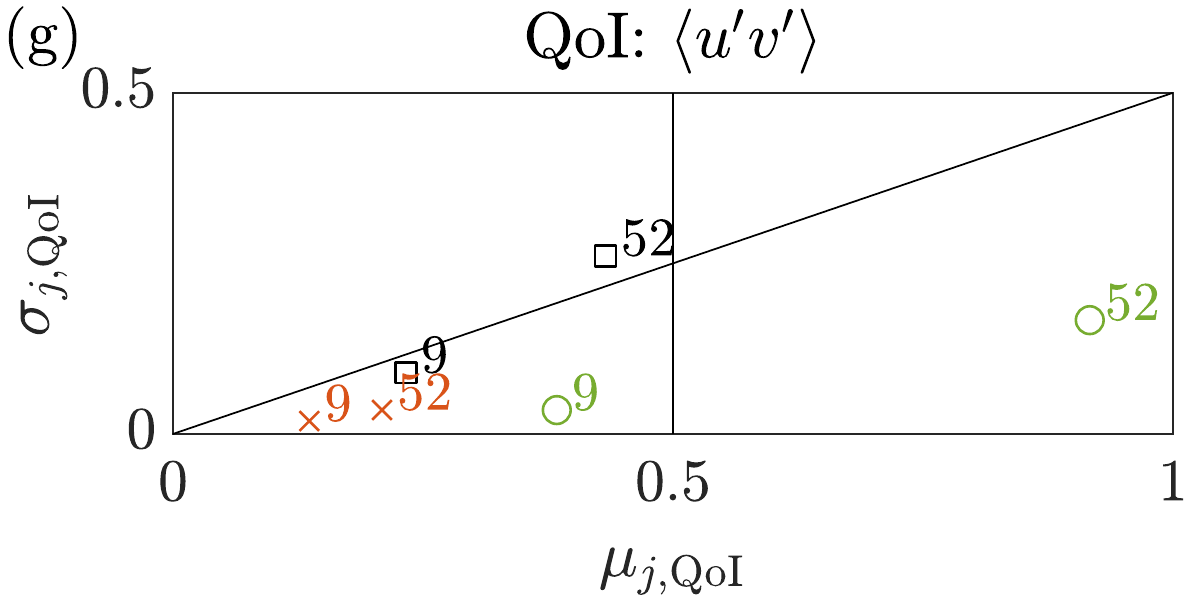}
    \includegraphics[height=0.162\textwidth]{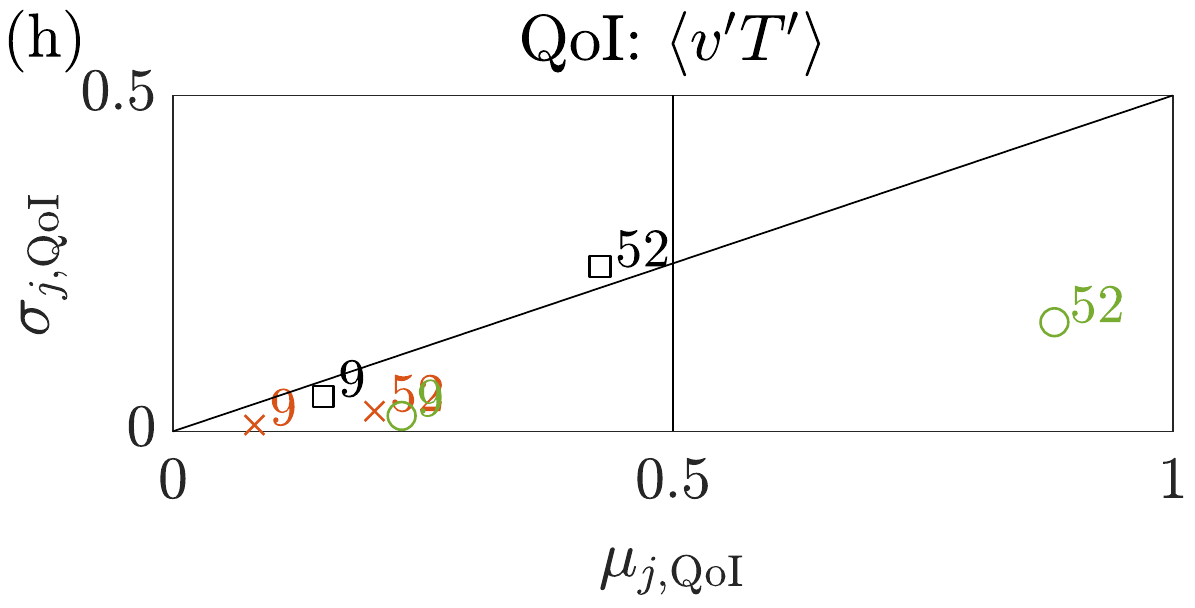}
    \includegraphics[height=0.162\textwidth]{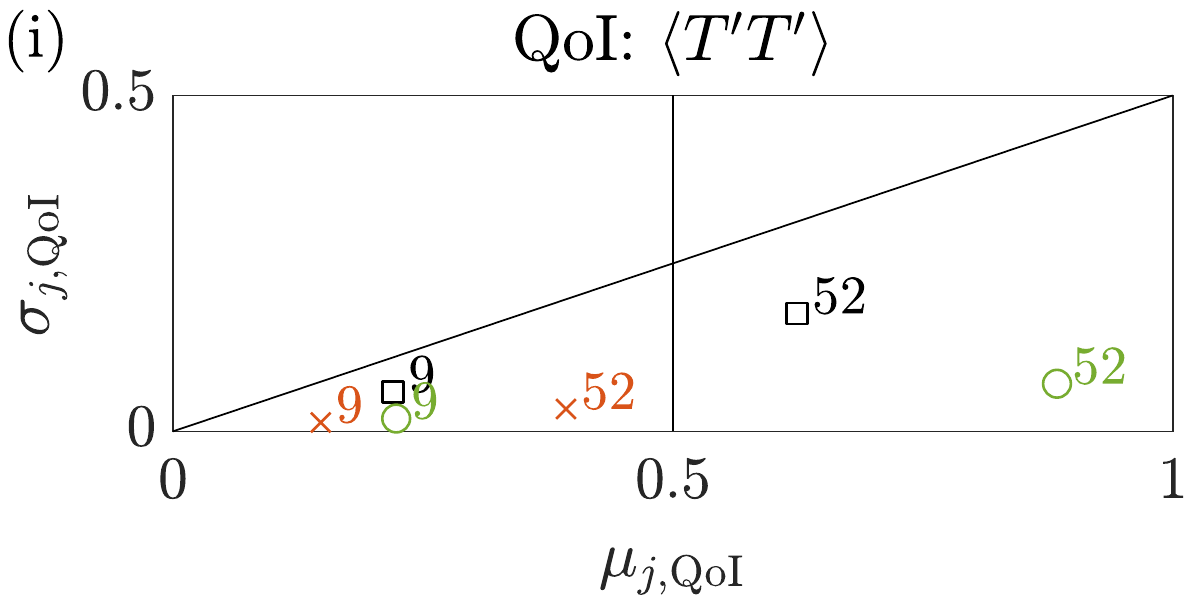}\\
    
    \caption{$\sigma-\mu$ diagram. 
    We use square symbols for un-grouped results, round symbols for group 1 results and cross symbols for group 2 results.}
    \label{fig:Results-S2-separate}
\end{figure}

We can verify our interpretation of the results by studying the perturbations' effectiveness and inconsistency for flows that remain at the early stage at $t=t_e$ and for flows that enter the late stage at an early time, i.e., $t<50$.
We refer to flows that remain at the early stage at $t=t_e$ as group 1 and flows that enter the late stage at a early time as group 2.
The cases are manually grouped.
We discard ambiguous cases like the one in figure \ref{fig:Results-S2-delta} (c).
We compute $\sigma$, $\mu$ for the group 1 cases and the group 2 cases as follows:
\begin{linenomath*}\begin{equation}
\begin{split}
    \mu_{j,{\rm QoI}, {\rm grp1}}=\frac{1}{N_{\rm grp 1}}\sum_\text{group 1 cases} e_{i,j,{\rm QoI}},~~~~~\sigma_{j,{\rm QoI},{\rm grp1}}^2=\frac{1}{N_{\rm grp1}}\sum_\text{group 1 cases} (e_{i,j,{\rm QoI}}-\mu_{j,{\rm QoI}})^2,\\
    \mu_{j,{\rm QoI}, {\rm grp2}}=\frac{1}{N_{\rm grp 2}}\sum_\text{group 2 cases} e_{i,j,{\rm QoI}},~~~~~\sigma_{j,{\rm QoI},{\rm grp2}}^2=\frac{1}{N_{\rm grp2}}\sum_\text{group 2 cases} (e_{i,j,{\rm QoI}}-\mu_{j,{\rm QoI}})^2,
\end{split}
\label{eq:sigma-mu-group}
\end{equation}\end{linenomath*}
where $N_{\rm grp1}$ and $N_{\rm grp2}$ are the number of cases in groups 1 and 2, respectively, and the subscripts grp1 and grp2 are for group 1 and group 2.
Figure \ref{fig:Results-S2-separate} shows the results.
For visualization purposes, we show results for perturbations 52 and 9, i.e., the most effective orientation perturbation and the most effective model coefficient perturbation for most QoIs.
The results for other perturbations are similar and are not shown here for brevity.
We see significantly lower $\sigma$ values for perturbations 52 and 9 when we consider cases in group 1 and group 2 separately.
This directly confirms our interpretation of the results in figure \ref{fig:Results-S3}.

In real-world applications, it is usually not difficult to know whether two models or two variants of the same baseline model are needed to accommodate different physics in the FCP space.
For example, a fluid mechanist who has the basic knowledge of shear layer's flow phenomenology knows that the physics is different in an early-stage shear layer and a late-stage shear layer and that different physics requires different models to accommodate.
The most notable example is the modeling of wall turbulence.
It is well known that the flow near a solid boundary behaves differently than flow away from a solid boundary.
In RANS, one accommodates this by incorporating the distance from the wall and a wall-treatment \cite{eisfeld2005advanced}.
When it comes to a stratified shear layer, one could incorporate, say, the local Richardson number and a low/high-Richardson-number treatment, which will be left for future investigation.
Accordingly, in practice, it can be problematic to include too many flows for model training, in which case, model training would needlessly trade off accuracy for robustness.

\subsection{Machine learning aided RANS modeling}
\label{sub:app-ml}

\begin{figure}
    \centering
    \includegraphics[width=0.48\textwidth]{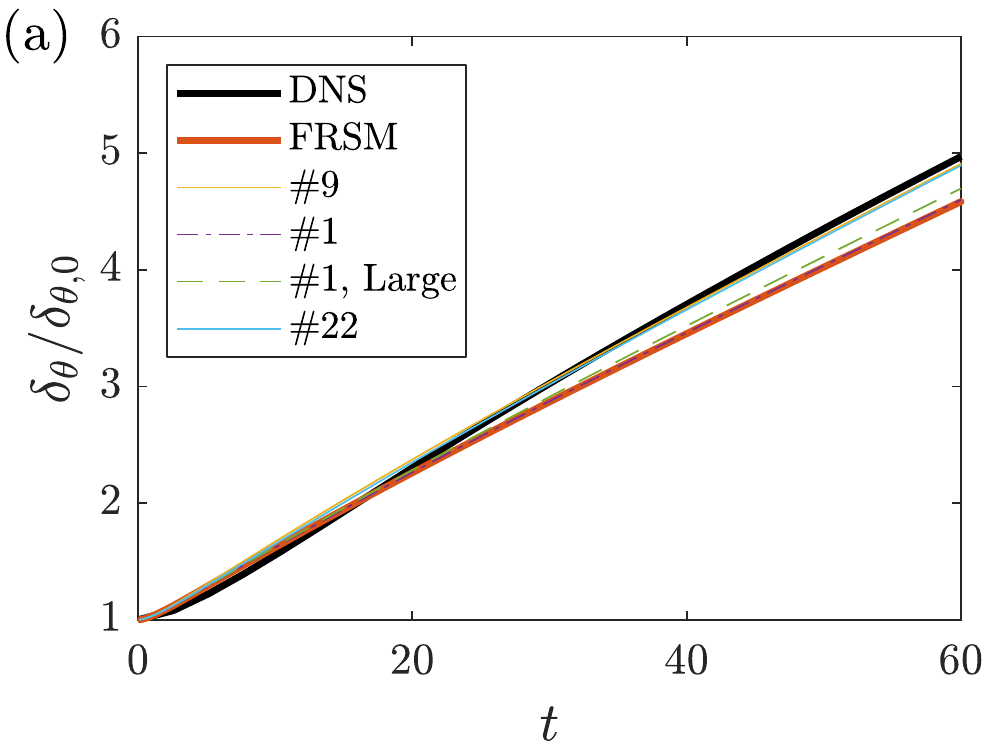}~~~~~
    \includegraphics[width=0.48\textwidth]{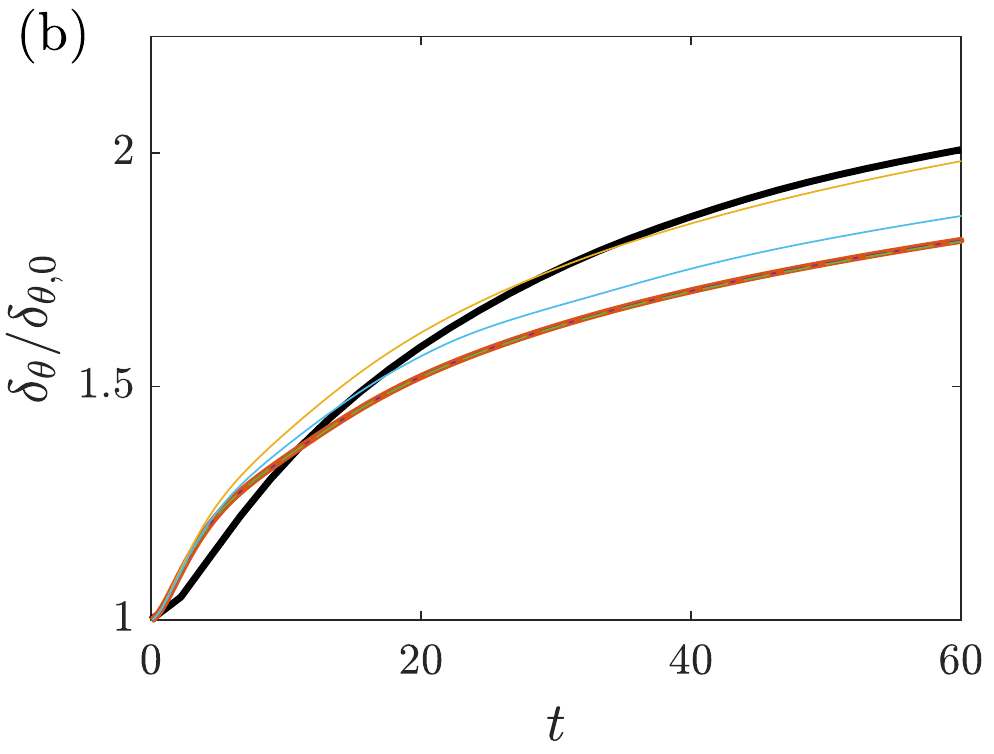}\\
    \includegraphics[width=0.48\textwidth]{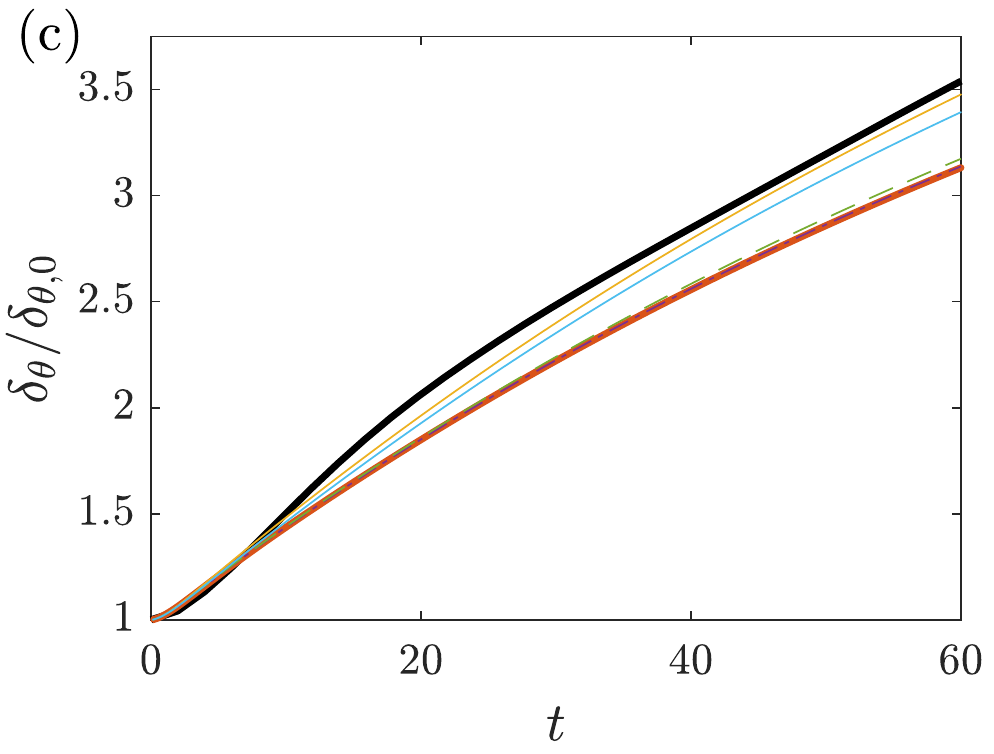}~~~~~
    \includegraphics[width=0.48\textwidth]{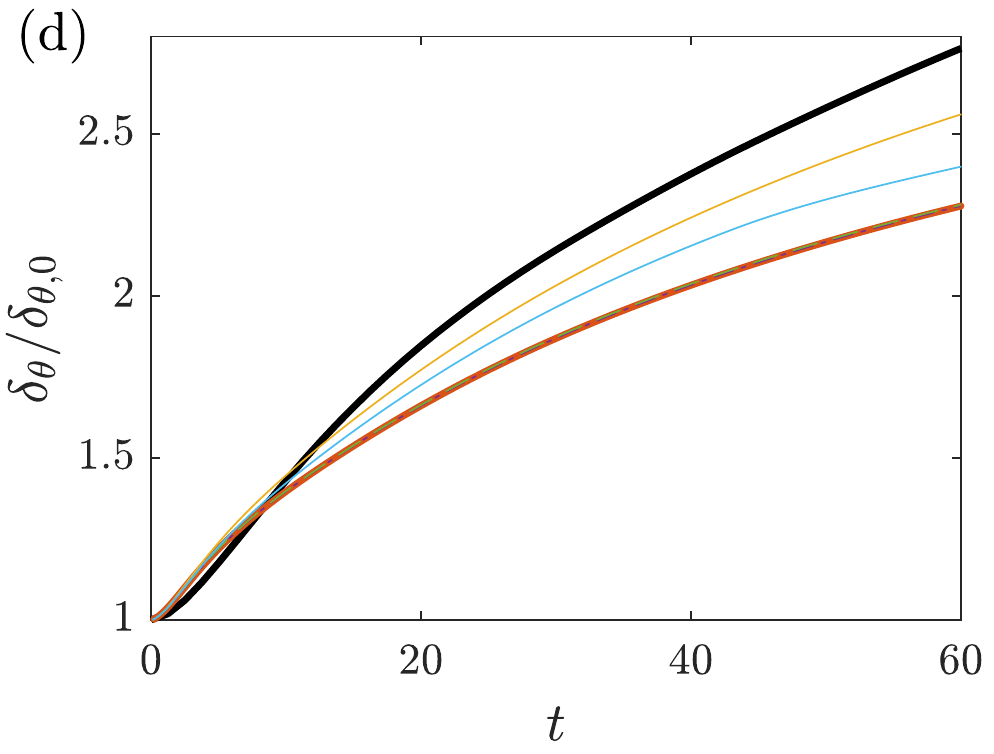}
    \caption{Shear-layer growth as a function of time.
    (a) the training condition, and (b,c,d) the test conditions. 
    Here, we calibrate the FRSM via perturbation 1, 9, and 22.
    Perturbation 1 is in the LE, LI quadrant.
    Perturbation 9 is in the HE, LI quadrant.
    Perturbation 22 is in the HE, HI quadrant.
    The legends are shown in (a).
    ``\#1 Large'' is where we vary $\sigma_\omega$ by 50\%.
    Perturbation \#1 does not have a notable impact on the QoI $\delta_\theta$, and ``\#1'', ``\#1, Large'', and ``FRSM'' collapse in (a).
    ``\#1'' and ``FRSM'' collapse in (b,c,d).
    }
    \label{fig:Results-modify}
\end{figure}

We have DNS data at one training condition, where the growth of the shear-layer thickness is shown in figure \ref{fig:Results-modify} (a).
We want to calibrate our FRSM to our data such that the calibrated model predicts shear layer growth more accurately at the training condition and at the other three test conditions, whose shear layer growth is shown in figure \ref{fig:Results-modify} (b, c, d).
To verify that the calibrated model does/does not work at the test condition, we will compare the FRSM result to DNS at the test condition.
Obviously, we do not use the DNS at the test condition for model calibration.
Again, the QoI is $\delta_\theta$.
We see from figure \ref{fig:Results-S3} (a) that perturbation 9 has high effectiveness and low inconsistency.
Calibrating the corresponding model coefficient $C_3$ to data would therefore lead to improvements that generalize.
On the other hand, perturbation 22 has high effectiveness and high inconsistency.
Calibrating the corresponding model term, i.e., the shape of the first term in Eq. \eqref{eq:Phiij}, would lead to improvements that work for the calibration data but do not generalize.
Although not shown in figure \ref{fig:Results-S3} (a), perturbation 1 has low effectiveness and low inconsistency.
Calibrating the corresponding model coefficient, $\sigma_\omega$, would not have a notable effect on the QoI.
Hence, from figure \ref{fig:Results-S3}, we conclude, {\rm a priori}, that calibrating $C_3$ would lead to an improvement at both the training condition and the test condition, calibrating the shape of the first term in Eq. \eqref{eq:Phiij} would lead to an improvement at the training condition but not necessarily at the test condition, and varying $\sigma_\omega$ would not lead to much improvement.
In the following, we calibrate $C_3$, $\sigma_\omega$, and the shape of the first term in Eq. \eqref{eq:Phiij} so that the model matches the data as closely as possible at the training condition and see if the calibrated model performs as expected at the other three test conditions.

Figure \ref{fig:Results-modify} shows the calibration results.
Figure \ref{fig:Results-modify} (a) shows the shear layer growth results at the training condition, and figure \ref{fig:Results-modify} (b, c, d) shows the shear layer growth results at the test condition.
We see that by calibrating $C_3$ such that the calibrated FRSM matches the DNS at the training condition, the calibrated FRSM model agrees more closely with the DNS at all test conditions.
We also see that matching the DNS at the training condition by calibrating the shape of the first term in Eq. \eqref{eq:Phiij}, the calibrated FRSM does not perform consistently at the three test condition.
Lastly, varying $\sigma_\omega$, even by 50\% (which is large perturbation), does not lead to a notable change in the shear layer growth. 
In all, the results confirm our expectation, and we can indeed determine {\it a priori} the applicable range of an improvement from global epistemic uncertainty quantification.

\section{Conclusions}
\label{sect:conclusion}

We develop a global epistemic UQ method that allows us to determine {\it a priori} a RANS model's applicable range, more specifically, to determine if a RANS model is able to extrapolate to a flow condition outside the training data.
The method consists of a local UQ method and a global screening method.
It evaluates a model perturbation (not necessarily a small perturbation) in terms of its effectiveness and consistency.
A perturbation that improves a RANS model's performance at one condition is going to generalize to other conditions in the flow controlling parameter space if and only if the perturbation has high effectiveness and low inconsistency.
We apply the method to FRSM of a stratified shear layer.
For this particular model, the model space has 55 dimensions; the FCP space is three dimensional; and 9 quantities of interest are considered.
A global epistemic UQ analysis of the FRSM model in this FCP space requires $O(10^3)$ RANS calculations.
The results point to perturbations of the production term, the pressure strain term, and the coefficients $C_\mu$, $C_3$ as the most effective perturbations among the 55 model perturbations for the 9 quantities of interest.
Following the global epistemic UQ results, we calibrate our FRSM to DNS at one flow condition by calibrating an effective and consistent model perturbation.
The calibrated model works well at the training condition.
More importantly, as expected from the global epistemic UQ analysis, the calibrated model generalizes to three other test conditions outside the training dataset.

\section*{Acknowledgement}
This work was sponsored by the US Office of Naval Research under contract N000142012315, with Dr. Peter Chang as Technical Monitor.
The DNSs are performed on ACI-ICS at Penn State.

\clearpage
\newpage

\begin{appendices}
\section{RANS details}
\label{app:RANS}

The RANS code we use is the in-house finite-volume unstructured solver NPHASE \cite{kunz2001unstructured}. 
The algorithm follows the established segregated pressure based methodology. 
We use a collocated variable arrangement and apply a lagged coefficient linearization  \cite{clift1994linear}. 
Continuity is imposed through pressure correction, based on the SIMPLE-C algorithm \cite{van1984enhancements}. 
In constructing cell face fluxes, a momentum interpolation scheme \cite{rhie1983numerical} is employed.

For the stratified shear layer calculations, we use a Cartesian mesh for a grid size of $N_x\times N_y\times N_z=4\times 401\times 2$. 
We refine the mesh near the centerline, using 28 grid points to resolve $2\delta_{\omega,0}$.
The flow is periodic in the streamwise direction, and we apply a symmetric condition at the side and top and the bottom boundaries.
The initial conditions are given as follows
\begin{equation}
\begin{split}
U &=-\frac{1}{2}\Delta U\tanh\left(\frac{2y}{\delta_{\omega,0}}\right),~ T=\frac{1}{2}\Delta T\tanh\left(\frac{2y}{\delta_{\omega,0}}\right), \\
\left<u'u'\right> &=\left<v'v'\right>=\left<w'w'\right> = \left[\Delta U ~{\rm TI}~\exp\left(-0.67 y^2/\delta_{\omega,0}^2\right)\right]^2, \\ 
\left<u'v'\right> &= -0.023C_{\mu}^{-1/4} \delta_{\omega,0}~\Delta U ~\frac{dU}{dy} ~ {\rm TI} \exp\left(-1.35y^2/\delta_{\omega,0}^2\right),\\
\left<u'w'\right>&=0,~ \left<v'w'\right>=0,\\ \left<u'\theta'\right> &=0,~ \left<v'\theta'\right>=0,~ \left<w'\theta'\right>=0,~ \left<\theta'\theta'\right>=0,
\end{split}
\end{equation}
where a Gaussian profile is used for the Reynolds stresses and the coefficients 0.67, 0.023, 1.35 are obtained by fitting to our DNS.
We integrate the RANS equations to $t_e=200$.

\section{DNS details}
\label{app:DNS}

We use the open-source finite-difference solver AFiD for our DNS. 
The code uses a second-order finite difference scheme and solves for fluid velocities on a staggered grid.
Time discretization uses a second-order Adams-Bashforth method. 
Further details of the solver can be found in Ref. \cite{van2015pencil} and the references cited therein \cite{ostilla2015multiple,kooij2018comparison,zhu2018transition}. 

For the stratified shear layer calculation, we apply a periodic condition in the streamwise and the spanwise directions and a free-slip condition at both the top and bottom boundaries.
We use a Cartesian mesh for the two DNSs. 
For the DNS with $Ri=0.1$, the domain size is $L_x\times L_y\times L_z=64.50\times 21.50\times 48.38$, and the grid count, $N_x\times N_y\times N_z= 384\times 128\times 512$. 
The other flow controlling parameters are $Pr=1.0$, $Re=1280$. 
For the DNS with $Ri=0.0$, the domain size is $L_x\times L_y\times L_z=64.50\times 21.50\times 32.25$, and the grid count $N_x\times N_y\times N_z= 384\times 128\times 192$.
The other flow controlling parameters are $Pr=1.0$, $Re=640$. 
Our grid resolution is the same as \cite{brucker2007evolution}.
The initial conditions are also the same as in Ref. \cite{brucker2007evolution}. 
Because the flow is transient, we run the same case 100 times for ensemble average.
\end{appendices}

\clearpage
\newpage

\bibliographystyle{ieeetr}
\bibliography{sample.bib}

\end{document}